\documentclass[letterpaper,twocolumn,english,aps,prb,superscriptaddress,letterpaper,twocolumn,citeautoscript,floatfix,showpacs,showkeys]{revtex4}
\usepackage[T1]{fontenc}
\usepackage[latin9]{inputenc}
\usepackage{color}
\usepackage{babel}
\usepackage{verbatim}
\usepackage{mathrsfs}
\usepackage{amsmath}
\usepackage{amssymb}
\usepackage{graphicx}
\PassOptionsToPackage{normalem}{ulem}
\usepackage{ulem}
\usepackage[unicode=true,pdfusetitle,
 bookmarks=true,bookmarksnumbered=false,bookmarksopen=false,
 breaklinks=true,pdfborder={0 0 0},backref=false,colorlinks=true]
 {hyperref}

\makeatletter

\pdfpageheight\paperheight
\pdfpagewidth\paperwidth

\newcommand{\lyxmathsym}[1]{\ifmmode\begingroup\def\b@ld{bold}
  \text{\ifx\math@version\b@ld\bfseries\fi#1}\endgroup\else#1\fi}

\usepackage{soul}\usepackage{mathrsfs}\usepackage{amsfonts}\usepackage{latexsym}\usepackage{array}\usepackage{tabularx}\@ifundefined{definecolor}
 {\usepackage{color}}{}
\usepackage{dcolumn}
\usepackage[normalem]{ulem}\usepackage{comment}\usepackage{bm}

\begin{document}

\title{Operation speed of polariton condensate switches gated by excitons }

\author{C. Ant\'{o}n}
\affiliation{Departamento de F\'{i}sica de Materiales, Universidad Aut\'{o}noma de Madrid, Madrid 28049, Spain}
\affiliation{Instituto de Ciencia de Materiales {}``Nicol\'{a}s Cabrera'', Universidad Aut\'{o}noma de Madrid, Madrid 28049, Spain}

\author{T. C .H. Liew}
\affiliation{School of Physical and Mathematical Sciences, Nanyang Technological University, 637371, Singapore}
 
\author{D. Sarkar}
\affiliation{Departamento de F\'{i}sica de Materiales, Universidad Aut\'{o}noma de Madrid, Madrid 28049, Spain}

\author{M. D. Mart\'{i}n}
\affiliation{Departamento de F\'{i}sica de Materiales, Universidad Aut\'{o}noma de Madrid, Madrid 28049, Spain}  
\affiliation{Instituto de Ciencia de Materiales {}``Nicol\'{a}s Cabrera'', Universidad Aut\'{o}noma de Madrid, Madrid 28049, Spain} 

\author{Z.~Hatzopoulos}
\affiliation{FORTH-IESL, P.O. Box 1385, 71110 Heraklion, Crete, Greece}
\affiliation{Department of Physics, University of Crete, 71003 Heraklion, Crete, Greece}

\author{P. S. Eldridge}
\affiliation{FORTH-IESL, P.O. Box 1385, 71110 Heraklion, Crete, Greece}

\author{P. G. Savvidis}
\affiliation{Department of Materials Science and Technology, Univ. of Crete, 71003 Heraklion, Crete, Greece}
\affiliation{FORTH-IESL, P.O. Box 1385, 71110 Heraklion, Crete, Greece}

\author{L. Vi{\~n}a}
\email{luis.vina@uam.es}
\affiliation{Departamento de F\'{i}sica de Materiales, Universidad Aut\'{o}noma de Madrid, Madrid 28049, Spain}
\affiliation{Instituto de Ciencia de Materiales {}``Nicol\'{a}s Cabrera'', Universidad Aut\'{o}noma de Madrid, Madrid 28049, Spain}
\affiliation{Instituto de F\'{i}sica de la Materia Condensada, Universidad Aut\'{o}noma de Madrid, Madrid 28049, Spain}

\begin{abstract}
We present a time-resolved photoluminescence (PL) study in real- and momentum-space of a polariton condensate switch in a quasi-one-dimensional semiconductor microcavity. The polariton flow across the ridge is gated by excitons inducing a barrier potential due to repulsive interactions. A study of the device operation dependence on the power of the pulsed gate beam obtains a satisfactory compromise for the \textit{on-off} signal ratio and -switching time of the order of~0.3 and $\thicksim50$~ps, respectively. The opposite transition is governed by the long-lived gate excitons, consequently the \textit{off-on}switching time is $\thicksim200$~ps, limiting the overall operation speed of the device to $\thicksim3$~GHz. The experimental results are compared to numerical simulations based on a generalized Gross-Pitaevskii equation, taking into account incoherent pumping, decay and energy relaxation within the condensate. 
\end{abstract}



\maketitle

\section{Introduction}

\label{sec:intro}

In recent years, the study of exciton-photon hybrid-particles, polaritons \cite{weisbuch_observation_1992}, is not only making great advances in terms of fundamental understanding but also in the quest for applications. Similarly to cold atoms, polaritons have a bosonic character allowing them to condense into a macroscopic coherent phase. Their comparatively lighter mass makes the condensation easier regarding the temperature \cite{kasprzak06nat}. Optical devices incorporating such condensates are promising candidates for new ultrafast, low-power consuming information processing components \cite{Cerna13ncomm,De-Giorgi:2012aa,Cancellieri14prl-starkShift}. Different structures, including one-dimensional ones, have been proposed recently for the realization of transistors \cite{johne10prb,Shelykh10prb-spin-transistor}, diodes \cite{Espinosa-Ortega13apl}, optical routers \cite{flayac13apl-polariton-router}, spin current controllers \cite{petrov13prb}, logic gates \cite{liew08prl_neurons} and for building a universal set of logic AND- and NOT-type gates \cite{Espinosa-Ortega:2013aa}.

Seminal experiments have established the foundations for the engineering of polariton components. These works include the injection of ultrafast polariton bullets \cite{amo09nat,adrados11prl-bullets}, the modulation of potential landscapes by optical means \cite{amo10prb}, the control of the spin \cite{martin02prl-polarization-control,amo10nphot-spin-switch,adrados11prl-bullets,malpuech06apl-polarization,grosso13arXiv-polariton-switch} as well as hysteresis effects due to bistability and multistability \cite{Paraiso:2010aa,gippius07prl-multistability,gavrilov13apl-multistability}. Photo-generated excitons allow the optical manipulation of the polariton flow and its amplification in micro-wires \cite{wertz10nphys-1D,Wertz12prl} and also in more complex structures such as a one-dimensional (1D) double-barrier tunneling diodes \cite{Nguyen:2013aa} or in a Mach-Zehnder interferometer \cite{Sturm:2014aa}.  

Recent experiments employ two non-resonant laser beams on 20-$\mu$m-wide ridges and have demonstrated the capability to block the polariton flow by optical means \cite{gao12prb-Polariton-transistor,Anton:2012aa,Anton:2013aa}. Different regions of the sample act as source (\emph{S}), gate (\emph{G}), and collector (\emph{C}) in close analogy to a transistor device {[}see Fig.~\ref{fig:scheme}(a){]}: a laser creates a polariton condensate at \emph{S} which serves as a source of polaritons, whose propagation is controlled by a weaker, pulsed beam at \emph{G}. There, photo-generated excitons induce a local potential barrier due to repulsive interaction with polaritons. In the \textit{on} state, the freely flowing polaritons feed a trapped condensate $\mathscr{C}_{C}$ close to the edge of the ridge at \emph{C}. In the case of the presence of a barrier at \emph{G,} polaritons are hindered to move towards \emph{C} and consequently they remain trapped temporarily in a condensate $\mathscr{C}_{S-G}$ between \emph{S} and \emph{G} (\textit{off} state). Recent time-resolved far- and near-field photoluminescence studies on this system report on the energy relaxation and dynamics of the polariton flow \cite{Anton:2012aa,Anton:2013aa}. 

In the present work, we investigate the dynamics of the device focusing on the operating conditions in terms of speed and \textit{on-off} signal contrast. In particular, the speed is mainly limited by the \textit{off-on} switching processes, which is conditioned by the lifetime of excitons at $G$. This paper is organized as follows. We discuss in Sec.~\ref{sec:sample_setup} the sample and the experimental setup. In Sec.~\ref{sec:exp}, we present and discuss our results. We first show (subSec.~\ref{subsec:charswitch}) the principle of operation, i.e. we discuss the switching dynamics for a selected gate power $P_{G}$. In Sec.~\ref{subsec:powoff}, we systematically investigate the dynamics of the device for varying $P_{G}$ in order to find an optimized operation point with an acceptable \textit{on-off} signal ratio. Finally, in Sec.~\ref{subsec:efflat}, we show that the actual two-dimensional (2D) character of the device does not affect adversely the operation for high and wide enough barriers at $G$, i.e., basically no significant amount of polaritons circumvents the gate. In Sec.~\ref{sec:model}, our experiments are compared with numerical simulations of the polariton condensate dynamics based on a generalized Gross-Pitaevskii equation, which is accordingly modified to account for incoherent pumping, decay and energy relaxation within the condensate. 

\section{Sample and experimental setup}

\label{sec:sample_setup}

\begin{figure}
\setlength{\abovecaptionskip}{-5pt} \setlength{\belowcaptionskip}{-2pt} 

\begin{centering}
\includegraphics[width=1\linewidth]{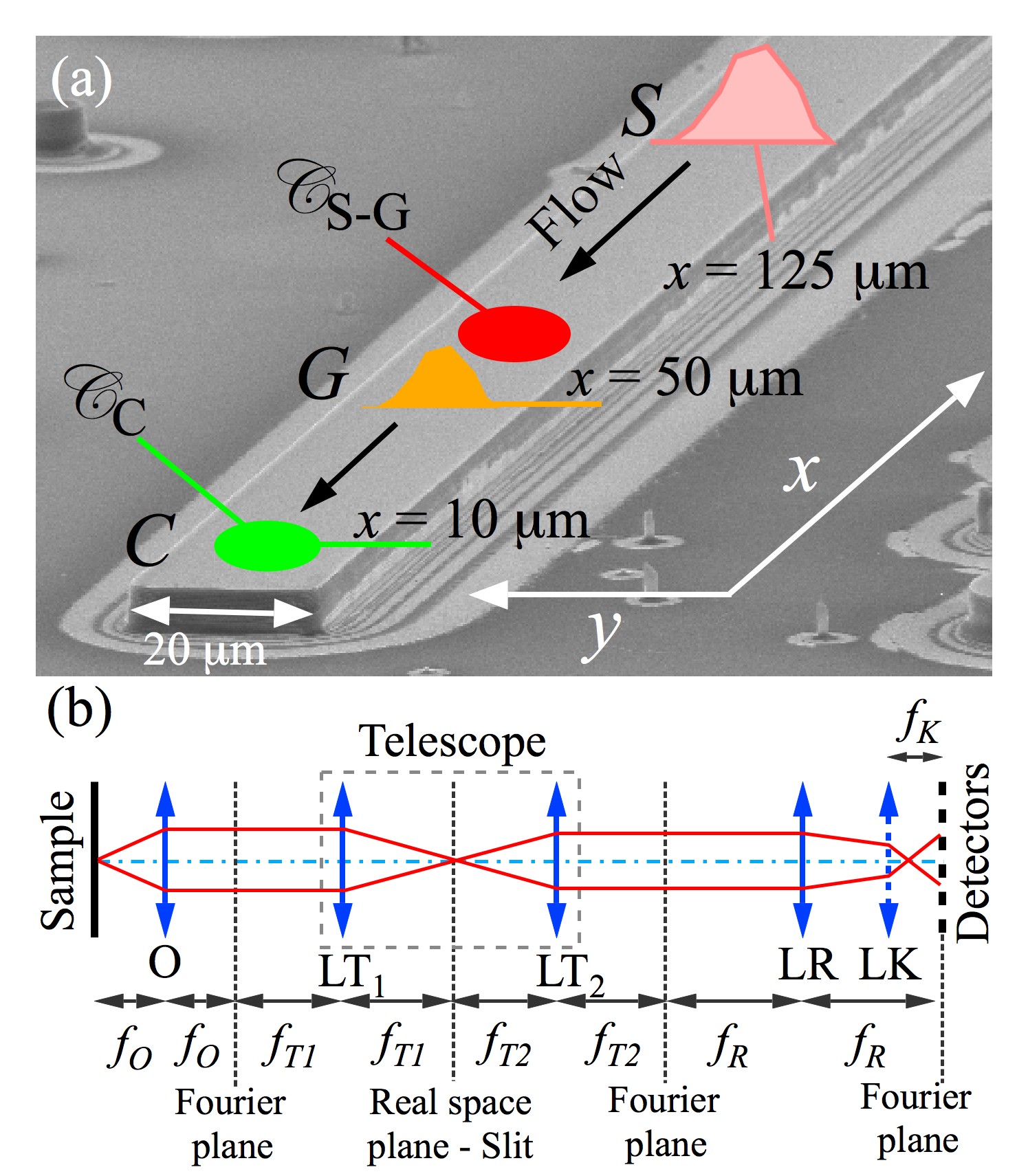} 
\par\end{centering}

\caption{(Color online) (a) Scanning electron microscopy image of a 20-$\mu$m wide ridge, including the excitation scheme with the continuous wave \emph{S} and pulsed \emph{G} beams, and the position of the trapped condensates: $\mathscr{C}_{S-G}$, between $S$ and G, and $\mathscr{C}_{C}$, at C. (b) Scheme of the imaging setup: The lenses \textquotedblleft{}LR\textquotedblright{} and \textquotedblleft{}LK\textquotedblright{} image the Fourier plane of the microscope objective \textquotedblleft{}O\textquotedblright{}. If lens \textquotedblleft{}LK\textquotedblright{} is removed, lens \textquotedblleft{}LR\textquotedblright{} images the real space. A slit is placed in the common focal plane of two lenses \textquotedblleft{}LT$_{1}$\textquotedblright{} and \textquotedblleft{}LT$_{2}$\textquotedblright{} with focal length $f_{T1}=80$ mm and $f_{T2}=150$ mm, respectively. $f_{O}{\color{red}{\normalcolor =20}}$ mm (numerical aperture = 0.3) , $f_{R}{\color{red}{\normalcolor =1000}}$ mm, and $f_{K}{\color{red}{\normalcolor =200}}$ mm are the focal lengths of the objective, \textquotedblleft{}LR,\textquotedblright{} and \textquotedblleft{}LK\textquotedblright{} lenses, respectively.}

\label{fig:scheme} 
\end{figure}

Figure~\ref{fig:scheme}(a) shows a scanning electron microscopy image of a 20-$\mu$m-wide ridge and illustrates the nomenclature we use, in close analogy to the terminology used in conventional electronic devices. We refer to different locations on the device as source $S$, gate $G$, and collector $C$ in accordance with the functionality of conventional transistor terminals. We assign correspondingly the symbols~$\mathscr{C}_{C}$ and~$\mathscr{C}_{S-G}$ to polariton condensates at and in-between these locations, respectively. The sample consists of a high-quality $5\lambda/2$ AlGaAs-based microcavity with 12 embedded quantum wells. Ridges with dimensions of $20\times300~\mu$m$^{2}$ have been fabricated by reactive ion etching, see Ref.~\onlinecite{Tsotsis12njp}. In our sample lateral confinement is insignificant as compared to much narrower 1D polariton wires \cite{tartakovskii98prb-1d,wertz10nphys-1D,Wertz12prl}. The chosen ridge is in a region of the sample corresponding to resonance, i.e. the detuning between the bare exciton and bare cavity mode is $\delta\backsim0$, and the Rabi splitting is $\Omega_{R}\sim9$ ~meV. The threshold power for condensation of polaritons under cw/pulsed excitation is $P_{th}^{cw}=7.5$~mW / $P_{th}=1.5$~mW.

Figure~\ref{fig:scheme}(b) shows a scheme of the experimental setup. The sample is mounted in a cold-finger cryostat and kept at~10~K. It is excited with continuous wave (cw) and pulsed Ti:Al$_{2}$O$_{3}$ lasers, both tuned to the first high-energy Bragg mode of the microcavity at 1.612~eV. The cw-laser acts as a source, and creates a continuous flow of polaritons towards both ends of the ridge. It is chopped at 300 Hz with an on/off ratio of 1:2 in order to prevent unwanted sample heating. The pulsed laser actuates as a gate at $G$ by means of 2~ps-long light pulses. The intensities and spatial positions of the $S$ and $G$ laser beams can be independently adjusted. We focus both beams on the sample through a microscope objective (O) to form two spots at $S$ and $G$ of 20~$\mu$m and 5~$\mu$m diameter, respectively. The distance between $S$ and $G$ ($C$) is $\backsim75$~$\mu$m ($\backsim115$~$\mu$m). Although a more compact device could be realized ($S$-$C$ distance $\backsim50$~$\mu$m), we have chosen larger separations between the terminals to clearly monitor the polariton dynamics, including the macroscopic propagation of the polariton flow, the trapping of polaritons between $S$ and $G$ ($\mathscr{C}_{S-G}$ formation) and the dynamics of the \textit{off-on} transition at $C$. The same objective is used to collect the photoluminescence (PL) within an angular range of $\pm18^{\circ}$. For momentum-space imaging \cite{Richard:2005aa}, an additional lens (LK) is placed in the optical path in order to image the Fourier plane of the microscope objective (O). We filter out the signal from $x>125$~$\mu$m, in order to facilitate the analysis of the momentum space images, placing a slit at the real-space image plane {[}Fig.~\ref{fig:scheme}(b){]}. The PL is analyzed with a spectrometer coupled to a streak camera obtaining energy-, time-, and spatial-resolved images, with resolutions of 0.4~meV, 15~ps, and 1~$\mu$m, respectively. Time $t=0$ corresponds to the pulse arrival at $G$. 

In our experiments polaritons propagate predominantly along the $x$-axis of the ridge, as demonstrated in Sec.~\ref{subsec:efflat}. The propagation in the $y$-direction, if any, is not crucial for the operation of our device. Therefore, all the images in the paper, where the $y$-direction is not shown, collect the PL along the $x$-axis from a $\Delta y=2$ $\mu$m-wide, central region of the ridge.
 
\section{Experimental results and discussion}

\label{sec:exp}

\subsection{Characterization of the device: polariton flow dynamics in real and momentum space }

\label{subsec:charswitch}

\begin{figure}
\setlength{\abovecaptionskip}{-5pt} \setlength{\belowcaptionskip}{-2pt} 

\begin{centering}
\includegraphics[width=1\linewidth]{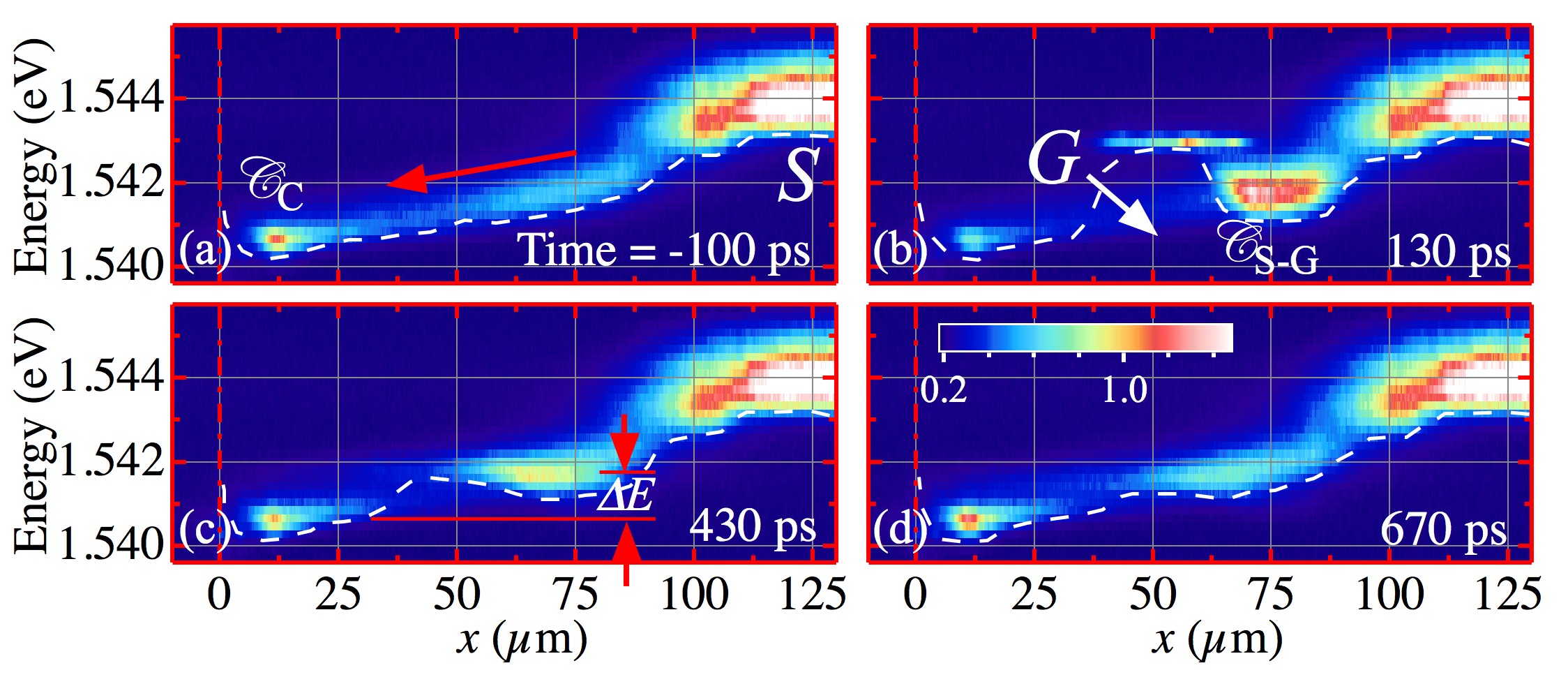} 
\par\end{centering}

\caption{(Color online) Energy vs. real space ($x$) at different times shown by the labels. \emph{S}, \emph{G}, $\mathscr{C}_{S-G}$, and $\mathscr{C}_{C}$ mark the source, the gate, the trapped condensate between \emph{S-G}, and the trapped one at \emph{C} positions, respectively. The white dashed lines show, as a guide to the eye, the effective potential experienced by polaritons, $V(x,t)$. The continuous flow of polaritons from \emph{S} to \emph{C} (in absence of \emph{G}) is indicated with a red arrow in panel (a). The relative blueshift $\Delta E$ between $\mathscr{C}_{S-G}$ and $\mathscr{C}_{C}$ is indicated with red arrows in panel (c). The intensity is coded in a linear false-color scale. A movie corresponding to this figure is provided in the Supplementary Material \cite{Note2}.}

\label{fig:spaceR} 
\end{figure}

We present a description of the polariton-flow switching dynamics
 in real- and moment um-space for a given power $P_{G}=0.55\times P_{th}$ . We show that the $\mathscr{C}_{C}$ emission intensity can be manipulated modulating temporarily the polariton flow by a potential barrier at $G$, which is induced by photo-generated excitons. The use of $P_{G}$ as a tuning parameter is of central interest in this work, and will be discussed in detail later in Sec. \ref{subsec:powoff}.  
  
Figure~\ref{fig:spaceR} shows the energy relaxation dynamics of the switching process versus real-space ($x$). The emission intensity is presented in false-color maps using a linear scale for different times. In Fig.~\ref{fig:spaceR}(a), at $t=-100$~ps, the PL of the polariton condensates $\mathscr{C}_{C}$ at $\sim10$~$\mu$m~($C$) and of the excitons located at $\sim125$~$\mu$m~($S$) are observed, respectively. The excitons at $S$, created by the cw-laser, act as source and emit at 1.544~eV. This energy is considerably blue-shifted with respect to the rest of the polaritons shown in the figure due to the high carrier density and repulsive interactions at this position. 

\begin{figure}
\setlength{\abovecaptionskip}{-5pt} \setlength{\belowcaptionskip}{-2pt} 

\begin{centering}
\includegraphics[width=1\linewidth]{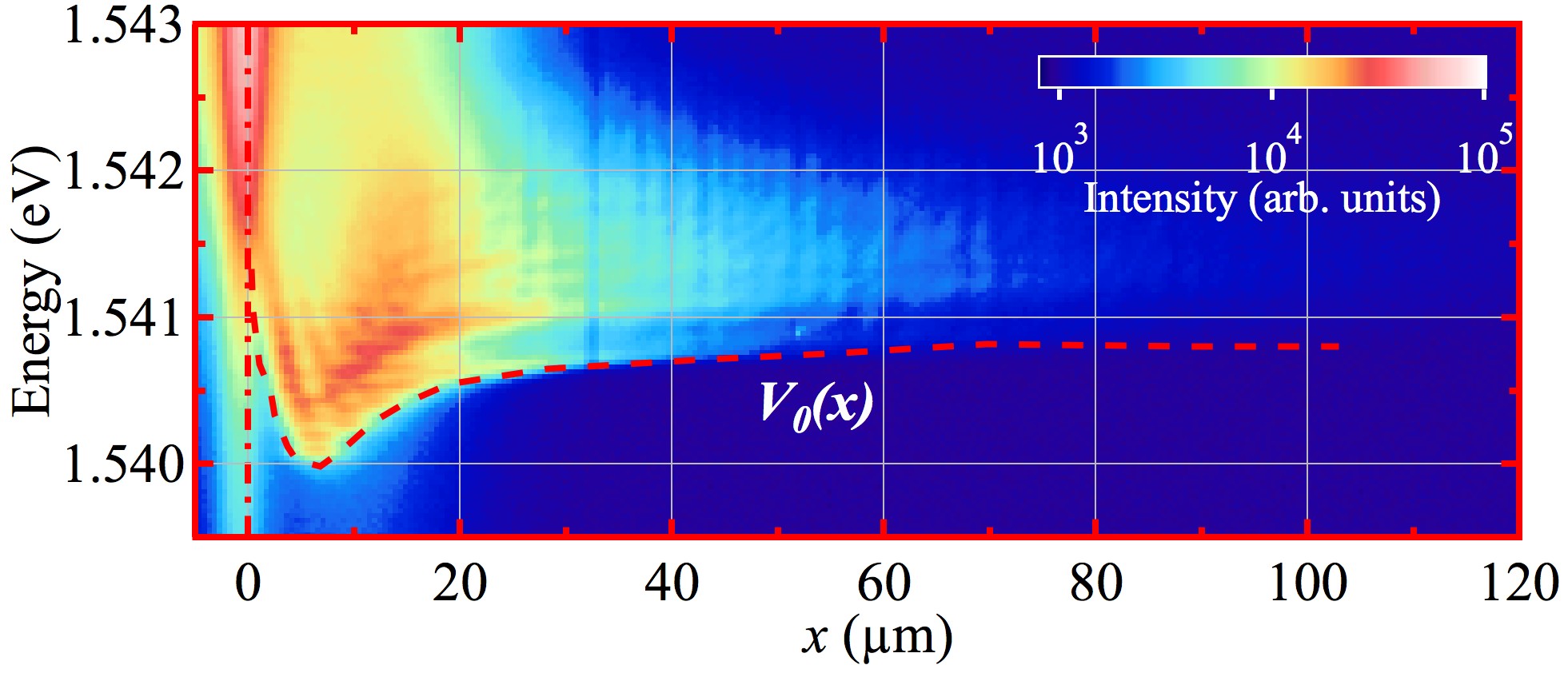} 
\par\end{centering}

\caption{(Color online) Polariton PL map as function of the emission energy and spatial position ($x$) under a pulsed, non-resonant (1.612 eV) and very weak (7 $\mu$W) excitation pump, placed at $x=10$ $\mu$m. The dashed red line shows, as a guide to the eye, the sample-geometry induced potential, $V_{0}(x)$. The intensity is encoded in a logarithmic, false-color scale.}

\label{fig:border-potential} 
\end{figure}

Dashed white lines in Fig.~\ref{fig:spaceR} outline the potential
landscapes. They are the sum of potentials due to sample geometry
and optically induced blue-shifts. It is experimentally impossible
to separate both contributions quantitatively, however, in order to
obtain an idea of the former potential, we create a dilute polariton
gas at the border of the ridge by a non-resonant and very weak pu mp at~$C$, see Fig.~\ref{fig:border-potential}. The potential trap, induced by the sample geometry, is visualized by the emission of the non-condensed polariton gas. The trap is located at a distance of $\sim7$~$\mu$m from the end of the ridge and it is $\sim1$~meV deep relative to the essentially flat potential away from the end of the device. By comparison with the results shown in Fig.~\ref{fig:spaceR}, we observe that the trap is shallower in the actual experiments. This is due to the strong blue-shift induced by the high density of the condensate at~$C$. Furthermore, we can conclude that the slope of potential along the ridge is almost entirely governed by the density of the polariton population and not by the sample geometry.

The potential slope causes polaritons to flow towards lower energies away from $S$ to the left and right (not shown) along the $x$-direction of the ridge. This flow is observed as a weak emission between $S$ and $C$. At $C$ the flow is stopped and polaritons accumulate at $\mathscr{C}_{C}$, emitting at 1.5407~eV. This energy is lower than that of the propagating polaritons as a result of a static potential, which has a minimum near the edge of the ridge \cite{Anton:2013ab}. $\mathscr{C}_{C}$ corresponds to the \textit{on} state of our polariton transistor device \cite{gao12prb-Polariton-transistor,Anton:2012aa,Anton:2013aa}. At $t=130$~ps, the barrier induced by the high carrier density created by the pulse at $G$ hinders the polariton flow towards $C$ {[}Fig.~\ref{fig:spaceR}(b){]}. At $G$, carriers emit at 1.543~eV. The polaritons accumulate before $G$~and form the condensate $\mathscr{C}_{S-G}$ emitting at an energy of 1.5408~eV, lying $\Delta E$=1.1 meV above $\mathscr{C}_{C}$. This configuration corresponds to the \textit{off} state of the device, since the polariton density at $C$ has decreased by $\sim40$~\%, compared to the situation shown in Fig.~\ref{fig:spaceR}(a). At $t = 430$~ps {[}Fig.~\ref{fig:spaceR}(c){]} the barrier height at $G$ lowers, since the carrier density has decreased, allowing polaritons again to move towards $C$, where the emission intensity increases again. Finally, at $t=670$~ps {[}Fig.~\ref{fig:spaceR}(d){]}, the device is basically back in the \textit{on} state. 

\begin{figure}
\setlength{\abovecaptionskip}{-5pt} \setlength{\belowcaptionskip}{-2pt} 

\begin{centering}
\includegraphics[width=1\linewidth]{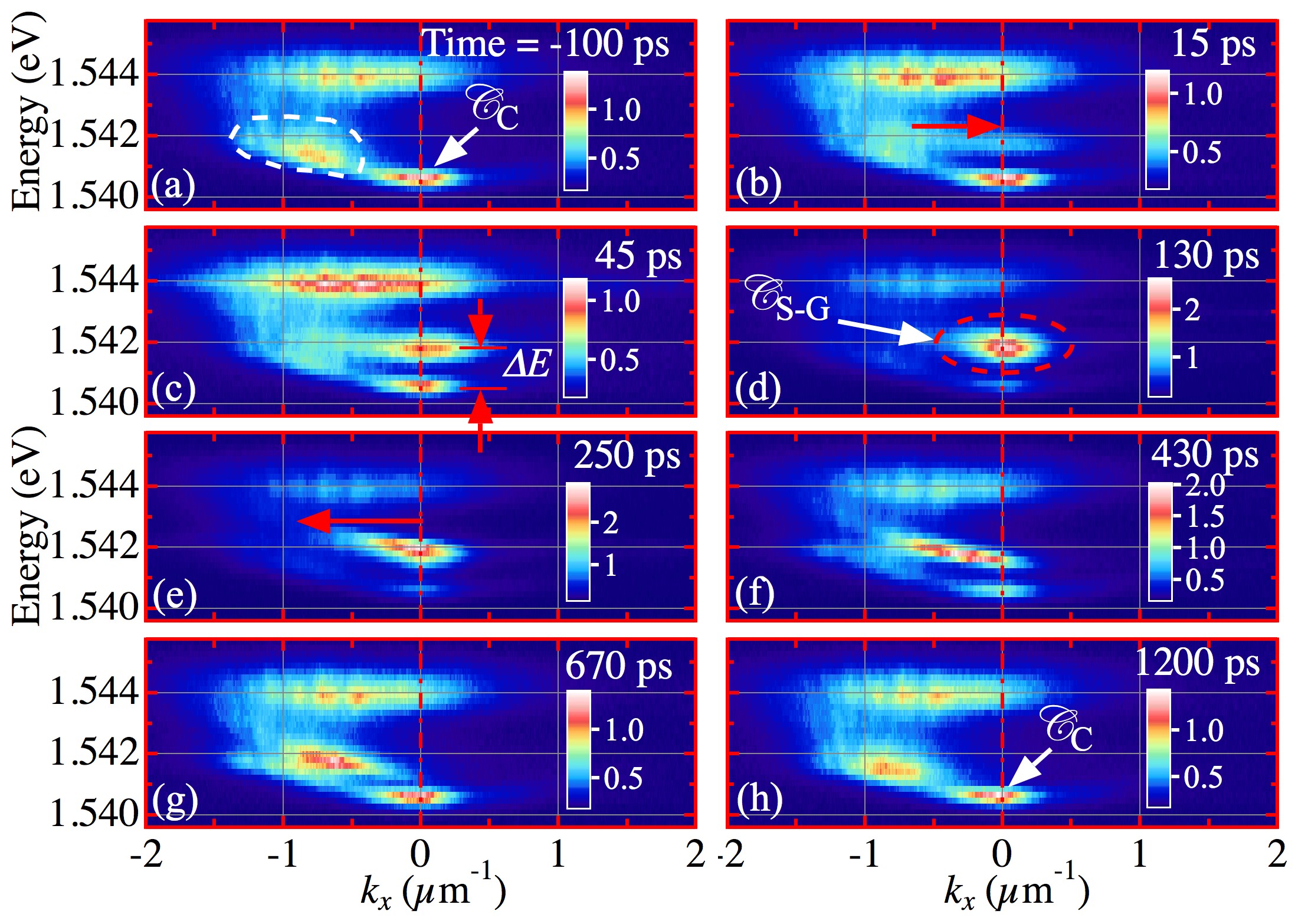} 
\par\end{centering}

\caption{(Color online) Energy vs. momentum space ($k_{x}$) at different times as shown by the labels. $\mathscr{C}_{S-G}$ and $\mathscr{C}_{C}$ mark the trapped condensates between \emph{S-G} and the trapped one at \emph{C} positions in real space, respectively. The acceleration of flowing polaritons is indicated by horizontal red arrows in panels (b) and (e). The relative blue-shift $\Delta E$ between $\mathscr{C}_{S-G}$ and $\mathscr{C}_{C}$ is indicated by vertical red arrows in panel (c). The intensity is coded in a linear, false-color scale. A movie corresponding to this figure is provided in the Supplementary Material \cite{Note2}.}

\label{fig:spaceK} 
\end{figure}

One can obtain a wealth of useful information on the polariton dynamics of the system from the real-space dynamics. However, an analysis of the momentum-space images complements and deepens the insight into the dynamics of the device operation. Figure~\ref{fig:spaceK} shows PL intensity maps versus emission energy and momentum along the $k_{x}$-direction. The PL intensity is coded in a linear false-color scale. Figure~\ref{fig:spaceK}(a) shows the PL at $t=-100$~ps. The flat and broad emission in momentum space, up to $k\thickapprox-1.5\:\mu$m$^{-1}$, at $E=1.544$~eV stems from excitons at $S$, as demonstrated in the analysis made previously in real-space. Below that energy, all polaritons appear to move mainly leftwards ($k_{x}<0$) towards $C$ (region enclosed by the white dashed line). Note that the right propagating flow from $S$ has been blocked, for the sake of clarity, as mentioned in Sec.~\ref{sec:sample_setup}. The peak energy and momentum of polaritons flowing towards $C$ is $1.541$~eV and $k_{x}=0.8\:\mu$m$^{-1}$, respectively. The polaritons slow down while approaching $C$, coming to a rest ($k_{x}=0$) at an energy $E=1.5405$~eV, where they form the trapped condensate $\mathscr{C}_{C}$. After the laser pulse has arrived at $G$, shown in Figs.~\ref{fig:spaceK}(b) and ~\ref{fig:spaceK}(c) at $t=15$~ps and $45$~ps, respectively, the polaritons decelerate (decrease of $\left|k_{x}\right|$). Concomitantly, an increasing PL is observed at $k_{x}=0$ and $E=1.542$~eV. At $t=130$~ps {[}Fig.~\ref{fig:spaceK}(d){]}, all the polaritons have been stopped and accumulate in the highly populated $\mathscr{C}_{S-G}$ (region enclosed by the red dashed line): this condition corresponds to the \textit{off} state of the switch. At longer times, the $\mathscr{C}_{S-G}$ polaritons start to accelerate towards $C$ again, as can be inferred from the shift of the PL peak from $k_{x}=0$ to more negative values {[}Figs.~\ref{fig:spaceK}(e) and \ref{fig:spaceK}(f){]}. The polaritons reach a maximum value of $k_{x}\thickapprox-1.3$~$\mu$m$^{-1}$ with a peak at $k_{x}=-0.8\:\mu$m$^{-1}$ {[}Fig.~\ref{fig:spaceK}(g){]}. Finally, Fig.~\ref{fig:spaceK}(h) shows that at $t=1200\:$ps, the same situation as shown in Fig.~\ref{fig:spaceK}(a) is encountered. The initial \textit{on} state is completely recovered, i.e., polaritons are able to flow again from $S$ to $C$; $\mathscr{C}_{S-G}$ disappears, and only the trapped $\mathscr{C}_{C}$ is present at $k_{x}=0$.

\begin{figure}
\setlength{\abovecaptionskip}{-5pt} \setlength{\belowcaptionskip}{-2pt} 

\begin{centering}
\includegraphics[width=0.7\linewidth]{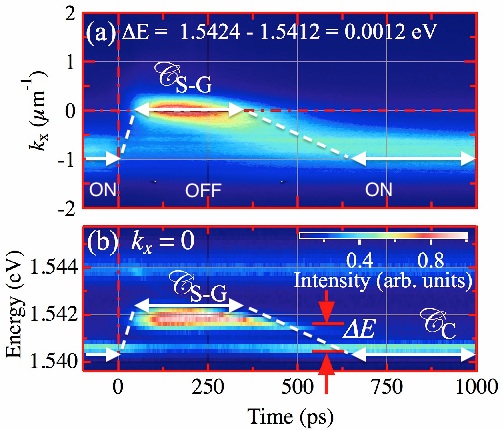} 
\par\end{centering}

\caption{(Color online) (a) Map of the polariton emission intensity, energy-integrated between 1.5412~and 1.5424~eV, versus $k_{x}$ and time. (b) Polariton emission intensity map for $k_{x}=0$ versus energy and time. $\mathscr{C}_{S-G}$ and $\mathscr{C}_{C}$ mark the trapped condensates between \emph{S }and \emph{G} as well as the trapped one at \emph{C}, respectively. Red arrows indicate the energy difference $\Delta E$ between $\mathscr{C}_{S-G}$ and $\mathscr{C}_{C}$. Both intensity maps are encoded in a normalized false-color scale shown in panel (b). The white arrows sketch the duration of the \textit{on} and \textit{off} states. The white dashed lines sketch the duration of the switching process. } 

\label{fig:spaceK-2} 
\end{figure}

Figure~\ref{fig:spaceK-2} completes the analysis of the switching dynamics in momentum space by paying attention to the flow in the $S-G$ region and $\mathscr{C}_{S-G}$. PL intensity maps are shown, which are encoded in a normalized, linear false-color scale. Figure~\ref{fig:spaceK-2}(a) shows a PL map, integrated between 1.5412 and 1.5424~eV, versus $k_{x}$ and time. The chosen range corresponds to the energy of the flow and $\mathscr{C}_{S-G}$. This figure clearly illustrates differences between the speed of the two switching processes. The \textit{off} state, which is characterized by PL from $\mathscr{C}_{S-G}$ at $k_{x}=0$, is reached within 100 ps after the gate pulse has arrived at $t=0$. In contrast, the \textit{on} state, which is distinguished by the flow with a peak momentum of $k_{x}\sim-0.8$~$\mu$m$^{-1}$, is recovered only after several hundreds of picoseconds. This asymmetry is a clear signature of the fast formation and slow decay dynamics of excitons. Figure~\ref{fig:spaceK-2}(b) provides a complementary perspective on the dynamics by showing the PL at $k_{x}=0$, i.e. that of both $\mathscr{C}_{C}$ and $\mathscr{C}_{S-G}$. The energy of the latter condensate is blue-shifted by $\Delta E\approx1$~meV with respect to that of $\mathscr{C}_{C}$ at $\sim1.5405$~eV. There are two remarkable effects on the dynamics of $\mathscr{C}_{S-G}$ and $\mathscr{C}_{C}$. First, there is a marked PL intensity drop of $\mathscr{C}_{C}$ during the \textit{off} state from $\sim50$ to $\sim400$~ps. Secondly, the temporal variation of the blueshift, which is induced by carrier density dependent polariton-polariton interactions, gives rise to an ``airfoil"-like shape of the $\mathscr{C}_{S-G}$ PL.

\subsection{Dependence of the off state on $P_{G}$}

\label{subsec:powoff}

The lateral width of the ridge is in principle a relevant parameter
that should be taken into account optimizing these 1D devices. Thinner ridges than the typical extension of the excitonic reservoir (> 500 $\mu$m$^{2}$) would ease the efficient blocking of the polariton flow and therefore improve the signal contrast between the \textit{on-off} states at $C$, but surface losses could be detrimental. Therefore, instead of varying the width of the ridge, we change the gate pump power $P_{G}$, which modifies the height and width of the gate barrier. This is in some degree equivalent to changing the width, while keeping the gate power constant. Also, in practical terms, it is more convenient to change the laser power than using different devices, where different potential landscapes would influence the results in an uncontrollable manner. 

After having demonstrated the working principle of the device in the previous subsection, we discuss now how to establish its optimal point of operation in the sense of a compromise between speed and an acceptable \textit{on-off} signal ratio. The speed of our device is mainly determined by the exciton dynamics at $G$. \cite{Note1} As shown in the previous section, this dynamics is of the order of hundreds of picoseconds, implying that the device cannot reach the THz range. As we will show here, weaker barriers allow a faster switching. However, the signal-ratio between both states diminishes, so that a trade-off between \textit{on-off} signal ratio and device speed has to be made. 

\begin{figure}
\setlength{\abovecaptionskip}{-5pt} \setlength{\belowcaptionskip}{-2pt} 

\begin{centering}
\includegraphics[width=1\linewidth]{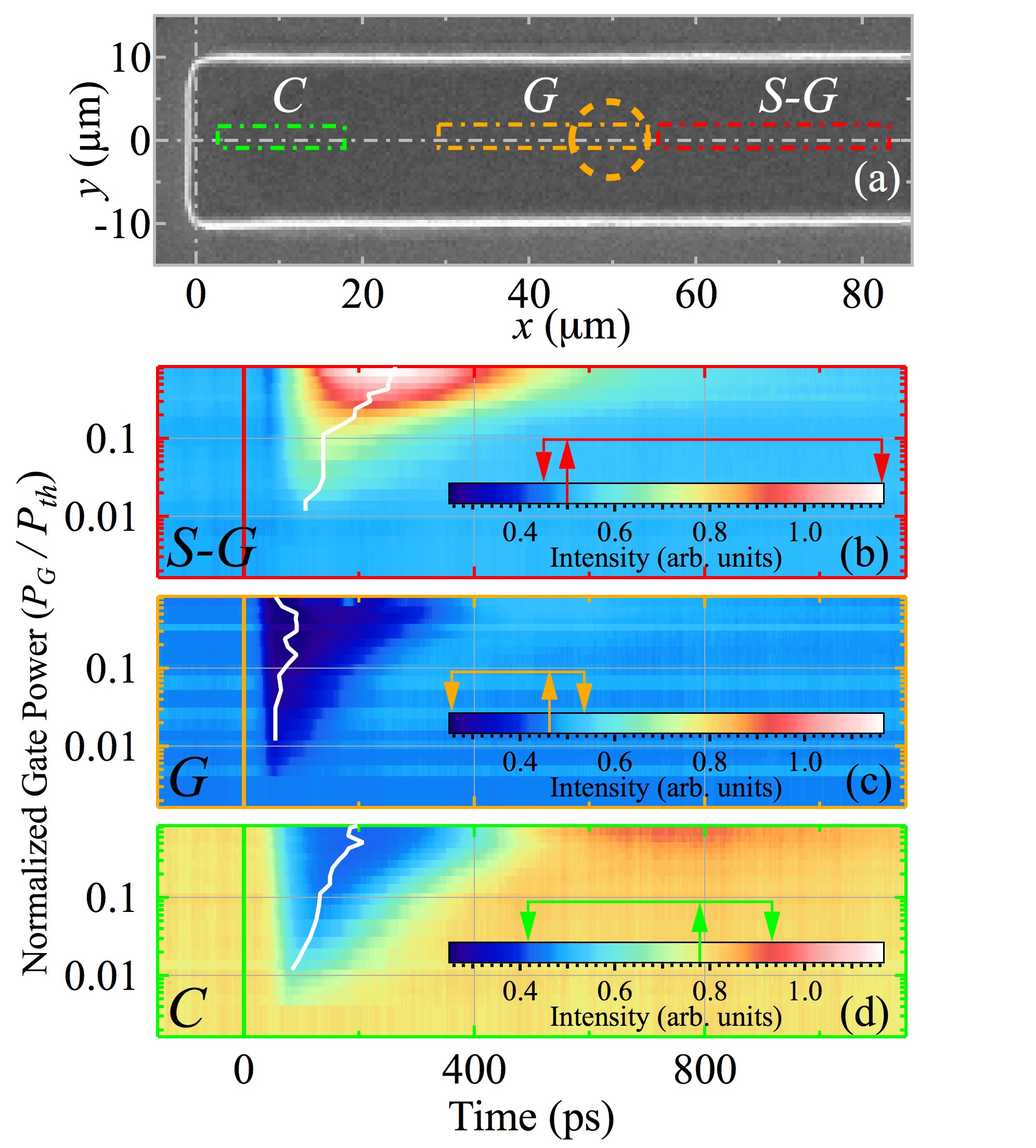} 
\par\end{centering}

\caption{(Color online)(a) Scanning electron microscopy image of a 20-$\mu$m wide ridge, marking in dot-dashed lines three selected spatial areas under study: flowing polaritons {[}red, ($S$-$G$){]}, the gate {[}orange, ($G$){]} and the collector {[}green, ($C$){]}. The position of \emph{G} is indicated with a dashed orange circle as a guide to the eye. (b) Polariton emission dynamics (spatially integrated in region $S$-$G$) as a function of $P_{G}$; for large values of $P_{G}$ polaritons are trapped forming the $\mathscr{C}_{S-G}$ condensate. (c) Polariton emission dynamics (spatially integrated in region \emph{G}) as a function of $P_{G}$. (d) Polariton emission dynamics (spatially integrated in region\emph{ C}) as a function of $P_{G}$; in this case the polariton emission corresponds to the $\mathscr{C}_{C}$ condensate. The intensity is coded in a linear false-color scale. The initial value of the intensity at each region is indicated with an up-pointing arrow at the color scale. The left/right down-pointing arrows mark the minimum/maximum PL intensities. The white lines give the times when the extreme PL intensity values are obtained as a function of $P_{G}$.} 

\label{fig:intensities} 
\end{figure}

In order to characterize the device operation, we present in Fig.~\ref{fig:intensities} a systematic study based on the PL intensity dynamics. This is analyzed at different regions as a function of the normalized gate power $P_{G}/P_{th}$. Figure~\ref{fig:intensities}(a) shows a top-view scanning electron microscope image of the part of the ridge relevant for the experiment. The dynamics in three centrally located rectangular regions, indicated by dot-dashed boxes, are investigated in detail. These regions correspond to the polariton flow against the gate ($S-G$), the gate ($G$) and the collector ($C$). Figures~\ref{fig:intensities}(b) - \ref{fig:intensities}(d) compile the PL dynamics versus $P_{G}$ (logarithmic ordinate) and time (linear abscissa) in each of the aforementioned regions. The up-pointing arrows on the linear false-color bars indicate the PL intensity at $t\leq0$; the left (right) down-pointing arrows mark the minimum (maximum) PL intensities. Figure~\ref{fig:intensities}(b) shows that for $P_{G}/P_{th}>0.01$, the PL increases after the gate pulse arrives, since the $G$ barrier blocks the polariton flow, resulting in the formation of $\mathscr{C}_{S-G}$. $\mathscr{C}_{S-G}$ persists longer for increasing $P_{G}$: for $P_{G}\sim P_{th}$, it lasts longer than 400 ps. The white line in Fig.~\ref{fig:intensities}(b) identifies the time when the maximum emission intensity of $\mathscr{C}_{S-G}$ is reached as a function of $P_{G}$. Note, that for powers $P_{G}>0.1\times P_{th}$ there is a small decrease of the PL at $t=45$~ps, which is due to thermal effects caused by the gate pulse \cite{amo10sst}.

Together with the formation of $ \mathscr{C}_{S-G}$, the blockade of the polariton flow is observed as a sudden decrease of the polariton population at $G$ and $C$, as shown in Figs.~\ref{fig:intensities}(c) and~\ref{fig:intensities}(d), respectively. The conspicuous PL drop observed in Fig.~\ref{fig:intensities}(c) is caused by the repulsive interactions between the photo-generated excitons and the passing polaritons at $G$. The time when the PL minimum intensity occurs is nearly independent of $P_{G}$, as shown by the white line in Fig.~\ref{fig:intensities}(c). In Fig.~\ref{fig:intensities}(d), the decrease of the emission intensity at $C$ is exclusively caused by the ceasing of the polariton flow. The best \textit{on-off} ratio obtained at $C$ is of the order of 50~\%. The temporal width of the drop at $C$ increases significantly with $P_{G}$, as evidenced by the blue region in Fig.~\ref{fig:intensities}(d), reaching a value of $\sim380$~ps at $P_{G}=0.8\: P_{th}$. It is worthwhile mentioning that the leading edge of the PL drop at $C$ appears at later times than that at $G$, due to the time of flow from remaining polaritons from $G$ to $C$.

\begin{figure}
\setlength{\abovecaptionskip}{-5pt} \setlength{\belowcaptionskip}{-2pt} 

\begin{centering}
\includegraphics[width=1\linewidth]{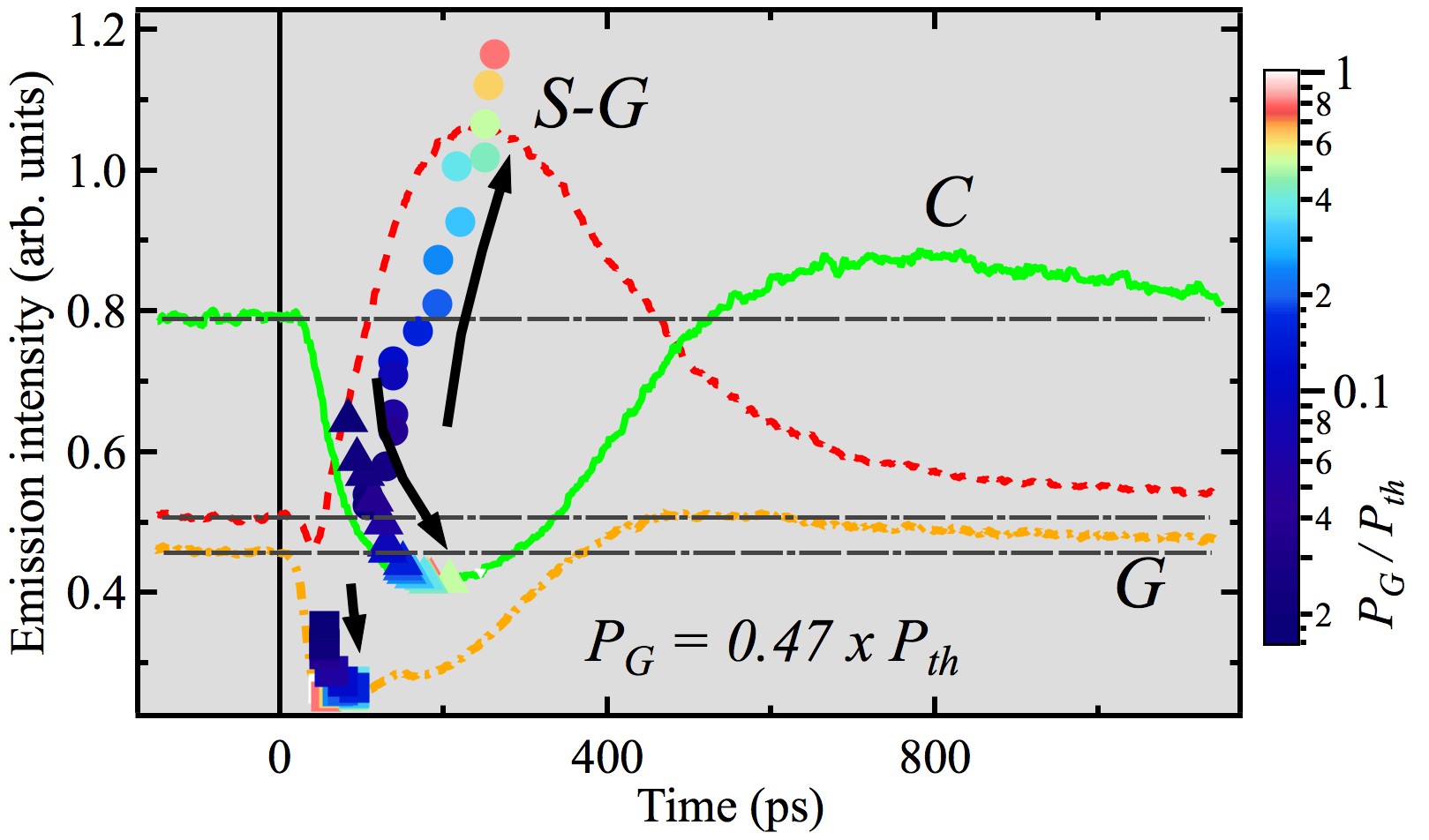}
\par\end{centering}

\caption{(Color online) Polariton intensity dynamics at $S$, $G$, $C$ regions in full red, orange and green colors, respectively, for $P_{G}=0.47\times P_{th}$. The initial value of the intensity is indicated with a dotted-dashed horizontal line. The scattered circles (squares, triangles), extracted from the white curves in Fig.~\ref{fig:intensities}, represent the maximum (minimum) PL values and their time of occurrence for the $S$-$G$ ($G$, $C$) regions. Their colors correspond to the $P_{G}$ value, encoded in a logarithmic, false-color scale shown at the right side of the figure.}

\label{fig:offdetail} 
\end{figure}

The optimal power $P_{G}$ for the best \textit{on-off} signal ratio is discussed in detail in Fig.~\ref{fig:offdetail}. The three curves correspond to horizontal cuts at $P_{G}/P_{th}=0.47$ in Figs.~\ref{fig:intensities}(b)-\ref{fig:intensities}(d) and show the emission intensity at the regions $S-G$, $G$ and $C$. In region S-G (dashed line), there is a small dip in intensity at $t=45$~ps as discussed previously. Thereafter, the emission intensity strongly increases and reaches its maximum at $t=250$~ps, where it doubles the initial value. The intensity at $G$ (dotted-dashed line) quickly decreases by $\sim50$~\% and recovers after a few hundred picoseconds. After recovering, it surpasses the initial value, due to the large polariton population released from the condensate $\mathscr{C}_{S-G}$. The higher value of the initial intensity at $C$ (full line) reflects the larger population in the collector as compared with the other regions of interest of the device. It drops down to $\sim50$~\% of its initial value at $\sim150$~ps. Again, as observed for $G$, there is an intensity overshoot due to the release of a large polariton flow from $\mathscr{C}_{S-G}$. Since the PL at $C$ does not drop down to zero, the \textit{on-off} signal contrast, $(I_{on}-I_{off})/(I_{on}+I_{off})$, is merely $\sim30$~\%. This contrast is slightly better at $G$ than at $C$ due to the fact that in the former case we have a potential hill, forcing polaritons away from $G$, while in the latter one there is a potential trap for the polaritons. 

The data presented as white curves in Fig.~\ref{fig:intensities} are shown now in Fig.~\ref{fig:offdetail} as scattered data points. The symbol color encodes $P_{G}/P_{th}$ in a logarithmic, false-color scale, as shown by the bar at the right of the figure. For the region $S-G$, the full circles indicate the maximum PL intensity (ordinate) and the time when it is reached (abscissa) for different $P_{G}/P_{th}$. Similarly, the squares and triangles show, for $G$ and $C$, respectively, the minimum PL intensity and the time when this is obtained. As aforementioned, in region $S-G$, the maximum PL intensity is increased and delayed with increasing $P_{G}$. The minimum intensity at $G$ drops down rapidly by $\sim50$ \% and then remains almost constant for $P_{G}/P_{th}\gtrsim0.2$. An inspection of the points corresponding to the collector reveals that the lowest emission intensities are reached also for the same $P_{G}$. Therefore, the optimum operation point, which gives the best compromise regarding contrast and speed is $P_{G}\sim0.2\: P_{th}$. 

Recently, different approaches have been investigated experimentally and theoretically to realize polariton switching systems in microcavities. These approaches involve various methods and features of polariton systems including parametric scattering\cite{Leyder2007prl} {[}power per gate/switch 9-40 mW, operation time 1 ns (theoretical){]}, hysteresis control\cite{De-Giorgi:2012aa} ($\sim30$ mW, 5 ps switching + 1 ns recovery), resonant blueshift\cite{Ballarini13natcomm} ($\sim5$ mW, 1 ns), spin domain walls\cite{amo10nphot-spin-switch,adrados11prl-bullets} {[}140 mW (pump)+ 4.5 mW (probe), $\sim70$ ps{]} and polariton condensate bullets\cite{Anton:2013ab} (44 mW, 150 ps switching + 250 ps recovery). Compared with these studies, our procedure lies within the lower range of gate pump powers of a few mW. 

Regarding the speed of the device, the switching into the \textit{off} state is faster than the reversing from \textit{off} to \textit{on} state. The speed of this reversal is certainly the main drawback of this device. It is determined by the long-lived exciton reservoir created at the gate in the \textit{off} state. Solutions would imply to find a way to make the excitons decay faster, or not making use of long-lived excitons at all. An \textit{on-off} transition time of a few picoseconds, but the reversal \textit{off-on} transition time still being in the hundreds of picoseconds, has been reported for resonant injection of polaritons in the lower polariton branch, avoiding the generation of excitons \cite{De-Giorgi:2012aa}. Ultrafast shifts of the lower and upper polariton branches exploiting the Stark effect in microcavities \cite{Hayat12prl-dynamic-stark-effect} have been proposed to implement optical switches with high repetition rates \cite{Cancellieri14prl-starkShift}. Furthermore, one can envision more complex geometries where, instead of excitons, polaritons flowing in a crosswise direction are employed to create the gate barrier. This work is a conceptual study of a polariton switch and this particular design may not provide an easy way for cascadability (capability to connect several devices in series) so far. However, there are other schemes, where the connectivity of propagating polariton signals has been recently demonstrated by a complex set of resonantly tuned lasers \cite{Ballarini13natcomm}.

\subsection{Leakage effects }

\label{subsec:efflat}

So far, for the sake of simplicity, we have considered one-dimensional dynamics of polaritons, analyzing the emission along the $x$-axis of a 2-$\mu$m wide, central stripe only. We have discussed the effect of the $G$ barrier height only, disregarding its lateral extension. In the following, we take into account that our system is not strictly one-dimensional, i.e. we also discuss the lateral effects depending on the size of $G$. For a given laser spot size at $G$, both the resulting extent and height of the excitonic barrier depend strongly on the power of the Gaussian laser beam. We show that the choice of working parameters is essential for a proper operation of the device, minimizing leakage currents around the gate. In order to observe the dynamics of these currents, it is convenient to avoid the steady state flow that is obtained under cw pumping at $S$. Therefore, we use now a pulsed laser with a power $P_{S}=4\: P_{th}$. 

\begin{figure}
\setlength{\abovecaptionskip}{-5pt} \setlength{\belowcaptionskip}{-2pt} 

\begin{centering}
\includegraphics[width=1\linewidth]{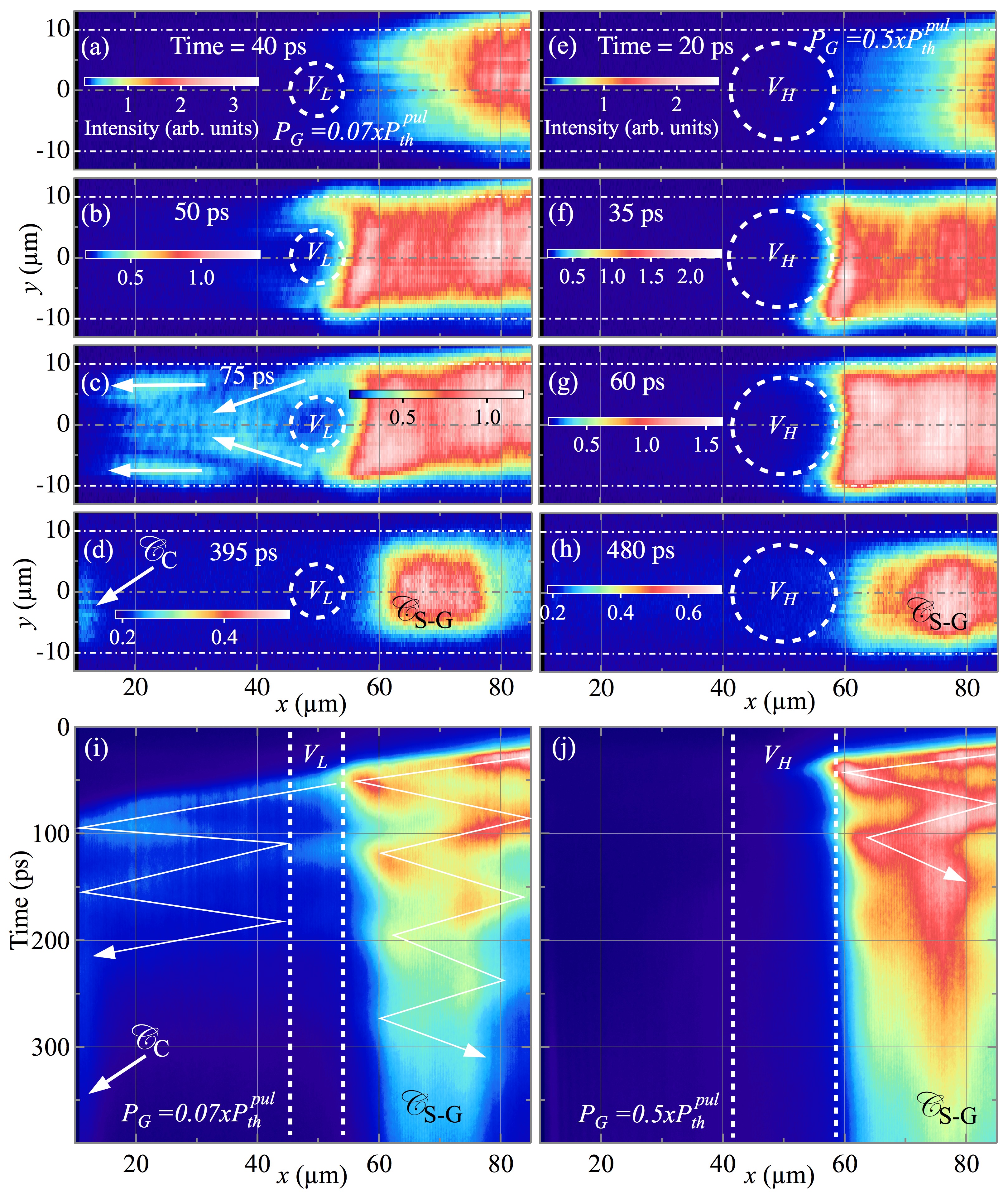} 
\par\end{centering}

\caption{(Color online) (a)-(h) Real-space emission intensity maps for different times as indicated by the labels. The intensity is encoded in a logarithmic false-color scale. $P_{S}=$ $4\times P_{th}$ and $P_{G}=$ $0.07\times P_{th}$ and $0.5\times P_{th}$ for the left and right column, respectively. A sketch of the extension of the barriers $V_{L}$ and $V_{H}$ is shown by dashed white circles. White arrows in panel (c) visualize the direction of the flow behind the barrier. (i), (j) Emission intensity maps versus time and the real-space $x$ direction. Real-space in the $y$ direction is integrated from $y=-10$ to $10~\mu$m. The widths of the potential barriers $V_{L}$ and $V_{H}$ are sketched by vertical, dashed lines. The trajectories of the polariton flow are indicated by white arrows.}

\label{fig:lateral} 
\end{figure}

Figure~\ref{fig:lateral} compiles the two-dimensional real-space dynamics of polariton propagation for two different gate powers $P_{G}$, keeping the laser spot size constant. The left and right columns show energy-integrated emission intensities, encoded in a logarithmic false-color scale, for $P_{G}=0.07\times P_{th}$ and $P_{G}=0.5\times P_{th}$, respectively. The former case creates a low and small barrier $V_{L}$, sketched by a dashed circle, which still permits polaritons to flow towards $C$. The latter case, which corresponds to the typical \textit{off} state discussed so far, induces a higher and larger barrier $V_{H}$, outlined by the bigger dashed circle.  

We discuss both situations in parallel. Initially, the front of the polariton flow approaches the barriers $V_{L}$ and $V_{H}$ at $x=50$, as shown in Figs.~\ref{fig:lateral}(a) and~\ref{fig:lateral}(e), respectively. Polaritons occupy the full width of the ridge, indicated by the dotted-dashed lines, and propagate uniformly with a well-defined velocity of $v_{x}=-1.8$~$\mu$m/ps, with negligible $k_{y}$ components. Figures~\ref{fig:lateral}(b) and~\ref{fig:lateral}(f) show instants when polaritons have collided against the $V_{L}$ and $V_{H}$ barriers, respectively. Since $V_{L}$ is significantly smaller than $V_{H}$, polaritons are able to overcome the $V_{L}$ barrier, basically bypassing it laterally, close to the edge of the ridge. In contrast, this is not possible in the case of $V_{H}$, where all polaritons remain blocked by the higher and wider barrier $V_{H}$. At times $75$ and 60~ps, shown in Figs.~\ref{fig:lateral}(c) and~\ref{fig:lateral}(g), respectively, polaritons continue to propagate. In Fig.~\ref{fig:lateral}(c), a complex structure is observed in the real-space density distribution of polaritons in the region from $x=40$ to $20$~$\mu$m. Here, polaritons, which bypass the barrier, propagate in negative $x$-direction but with a small $k_{y}$ component, sketched by the slanted arrows. Further apart from $G$, there are two lobes of polaritons with ${\color{red}{\normalcolor k_{y}\sim0}}$, indicated by horizontal arrows. Figures~\ref{fig:lateral}(d) shows that at longer times polaritons do not bypass the $V_{L}$ barrier anymore: the fast energy relaxation of polaritons as compared with the decay rate of $V_{L}$ results in the trapping of $\mathscr{C}_{S-G}$. Furthermore, the weak emission at $x\sim10$~$\mu$m arises from $\mathscr{C}_{C}$, formed by polaritons which have formerly overcome $V_{L}$. The trapping of $\mathscr{C}_{S-G}$ at very long times is also clearly observed in Fig.~\ref{fig:lateral}(h). 

Figures~\ref{fig:lateral}(i) and~\ref{fig:lateral}(j) show the real-space dynamics in $x$-direction as PL intensity maps encoded in a logarithmic false-color scale for the two $P_{G}$ values, respectively. In the $y$-direction the full lateral width of the ridge is integrated. The widths of the potential barriers $V_{L}$ and $V_{H}$ are sketched by vertical, dashed lines. The trajectories of the polariton flow are indicated by white arrows. Figure~\ref{fig:lateral}(i) illustrates that polaritons, having bypassed the barrier, experience a zig-zag movement on the left side of $V_{L}$ in the time range from $\sim50$ to $\sim200$~ps. They are reflected between the barrier and the potential wall at the end of the ridge \cite{Wertz12prl,Anton:2013ab}, where, finally at $\sim250$~ps, a part of these polaritons is trapped and form $\mathscr{C}_{C}$. The main part of the initial polariton flow remains trapped between $S$ and $G$, also moving in a zig-zag path. The polaritons gradually lose their kinetic energy, which is evidenced in the figure by the changing slope of the white lines. This is caused by a decrease of the potential gradient due to a reduced blue-shift originating from a falling carrier population. In principle, this could affect the device operation speed. However, this effect can be neglected in our device since the limiting factor is the existence of long-lived excitons at the gate. Eventually at $\sim250$~ps, the stationary $\mathscr{C}_{S-G}$ forms at $\sim70$~$\mu$m. Figure~\ref{fig:lateral}(j) shows the case of the higher and wider barrier $V_{H}$. The high exciton density at $G$ forms a $\sim2$0~$\mu$m-wide barrier potential. It completely blocks and reflects the polariton flow, which rapidly loses its kinetic energy. Finally at $\sim100$~ps, $\mathscr{C}_{S-G}$ is formed at $\sim80$~$\mu$m. Movies corresponding to Fig.~\ref{fig:lateral}
are provided in the Supplementary Material \cite{Note2}. In those movies, for times around 60 ps, the just discussed oscillations in the propagation of polaritons between $S$ and $G$ are clearly observed.

We have shown here the influence of $P_{G}$ on the existence of leakage currents around the $G$ barrier, which implies the worsening of the \textit{on-off} ratio of the device. An alternative approach to avoid leakage could be either a geometrical constriction at $G$ and/or the use of an elongated profile, along $y$, for the laser spot at $G$.

\section{Model}

\label{sec:model}

\subsection{Simulations of the experimental results}

In this section, we review the theoretical model used to simulate our experimental results, based on a phenomenological treatment of polariton energy-relaxation processes, which we have previously used to study transistor dynamics and propagation under a pulsed source \cite{Anton:2013aa,Anton:2013ab}. It is important to note that energy relaxation occurs in multiple stages in our experiment. First, the non resonant pump creates a reservoir of hot excitons, which can relax in energy to form polaritons. Given that excitons diffuse very slowly and that energy relaxation is local, these polaritons must form initially at the same position as the pump source. The polaritons then travel down a potential gradient, which is caused by repulsion between polaritons and hot excitons. It is the ability of polaritons to further relax their energy as they travel that allows them to be so sensitive to a changing potential landscape, as seen in a variety of different experiments in planar microcavities \cite{Balili2007science,Cristofolini2013prl}, one-dimensional microwires \cite{wertz10nphys-1D,Tanese2013ncomms} and in condensate transistors \cite{gao12prb-Polariton-transistor,Anton:2012aa}. 

The polaritons, having condensed from excitons, are known to be coherent \cite{kasprzak06nat}, which allows their description in terms of a Gross-Pitaevskii type equation \cite{carusotto04prl} for the polariton wave-function, $\psi(x,t)$: 

\begin{align} i\hbar\frac{d\psi(x,t)}{dt}&=\left[\hat{E}_{LP}+\alpha|\psi(x,t)|^2+V(x,t)\right.\notag\\ &\left.+i\hbar\left( rN_A(x,t)-\frac{\Gamma}{2}\right)\right]\psi(x,t)+i\hbar\mathfrak{R}\left[\psi(x,t)\right]\label{eq:GP} 
\end{align}
Here, $\hat{E}_{LP}$ represents the kinetic energy dispersion of polaritons, which at small wavevectors can be approximated as $\hat{E}_{LP}=-\hat{\nabla^2}/\left(2m\right)$ with $m$ the polariton effective mass. $\alpha$ represents the strength of polariton-polariton interactions. $V(x,t)$ represents the effective potential acting on polaritons~\cite{wouters07pra} and can allow or block the polariton propagation depending on its shape. $V(x,t)$ can be divided into a contribution from three different types of hot exciton states, which will be described shortly, as well as a static contribution due to the wire structural potential, $V_0(x)$: 
\begin{equation} 
V(x,t)=\hbar g\left[N_A(x,t)+N_I(x,t)+N_D(x,t)\right]+V_0(x) 
\end{equation} 
where subindices $A$, $I$ , $D$ refer to active, inactive and dark excitons, respectively. Experimental characterization has revealed that the static potential, $V_0(x)$, is non-uniform along the wire and exhibits a dip in the potential near the wire edge \cite{Anton:2012aa,Anton:2013aa}. The terms $rN_A(x,t)$ and $\Gamma/2$ appearing in Eq.~(\ref{eq:GP}) represent modifications to the Gross-Pitaevskii equation that describe gain and loss in the system \cite{wouters07pra,keeling_spontaneous_2008}, caused by the condensation of polaritons from the exciton reservoir $N_A(x,t)$ and polariton decay $\Gamma$, respectively.

\begin{figure}[t]
\setlength{\abovecaptionskip}{-5pt} \setlength{\belowcaptionskip}{-2pt} 

\begin{centering}
\includegraphics[width=1\linewidth]{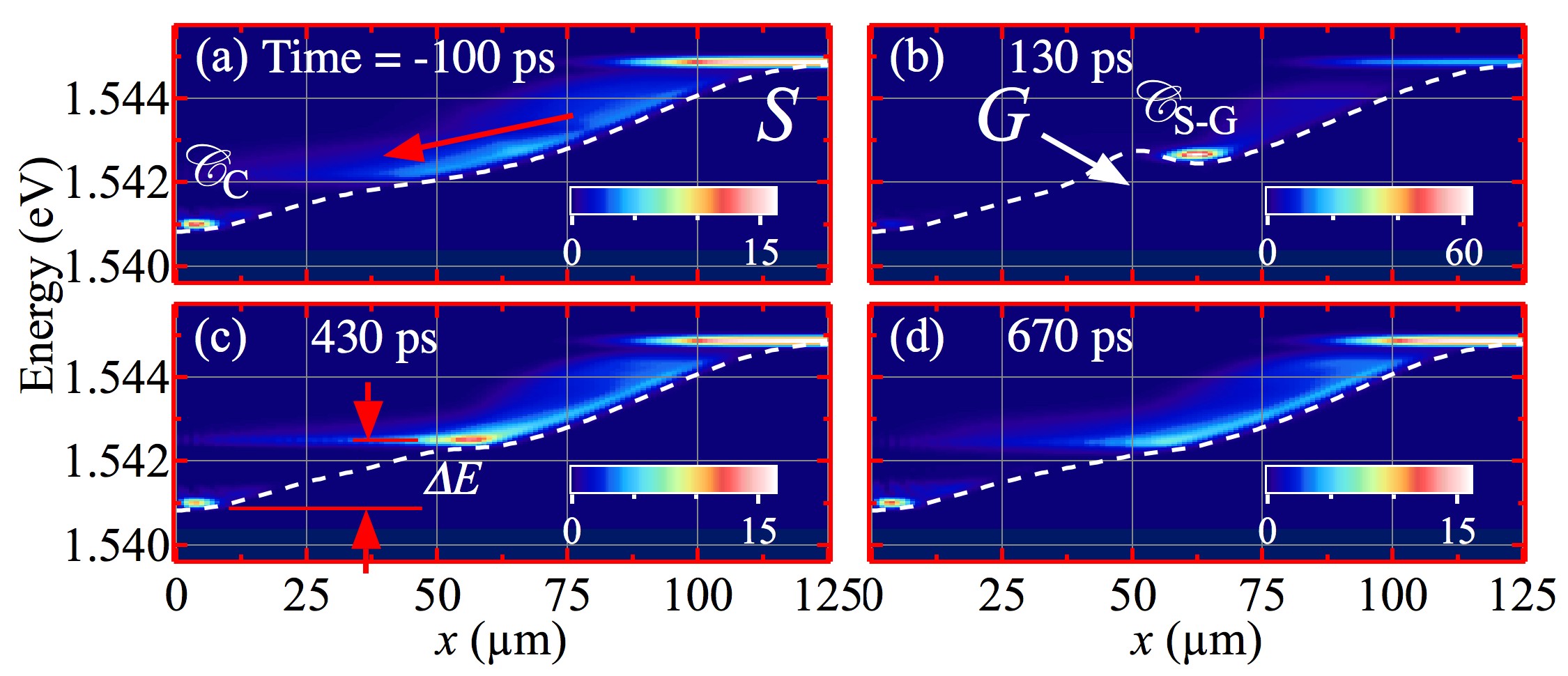} 
\par\end{centering}

\caption{(Color online) (a)-(h)\textcolor{red}{{} }Calculated emission intensity false-color scale maps vs. energy and real space ($x$) for different times shown by the labels. $\mathscr{C}_{S-G}$ and $\mathscr{C}_{C}$ mark the trapped condensates between \emph{S-G} and the trapped one at \emph{C} positions in real space, respectively.The relative blue-shift $\Delta E$ between $\mathscr{C}_{S-G}$ and $\mathscr{C}_{C}$ is indicated by vertical red arrows in panel (c). }

\label{fig:TheoryEx} 
\end{figure}

Previous studies of polariton dynamics have revealed that not all excitons are available for direct scattering into the polariton states \cite{Wouters2009prb}. Rather, one can distinguish between ``active'' excitons and ``inactive'' excitons \cite{lagoudakis11prlSHORT,Manni2012natcomm}. The active excitons have the correct energy and momentum for direct stimulated scattering into the condensate [and so appear as the incoherent gain term in Eq.~(\ref{eq:GP}) with $r$ the condensation rate]. However, non-resonant pumping creates initially excitons with very high energy that must first relax before becoming active. We thus identify the ``inactive'' reservoir that feeds the active reservoir. In principle, inactive excitons can also relax into dark exciton states that are uncoupled to polaritons. The three exciton densities ($N_A$, $N_I$, and $N_D$) are in general spatially and time dependent. They each give a repulsive contribution to the effective polariton potential with strength described by the parameter $g$. The dynamics of the exciton densities is described by rate equations \cite{Anton:2013aa}:
\begin{align} \frac{dN_A(x,t)}{dt}&=-\left(\Gamma_A+r|\psi(x,t)|^2\right)N_A(x,t)+t_RN_I(x,t),\label{eq:NA}\\ \frac{dN_I(x,t)}{dt}&=P(x,t)-\left(\Gamma_I+t_R+t_D\right)N_I(x,t),\label{eq:NI}\\ \frac{dN_D(x,t)}{dt}&=t_DN_I(x,t)-\Gamma_DN_D(x,t).\label{eq:ND} 
\end{align} 
The constants $t_R$ and $t_D$ describe the transfer of inactive excitons into lower energy active and dark exciton states, respectively. We neglect any nonlinear conversion between bright and dark excitons\cite{vina2004sst,Shelykh2005ssc,Anton:2013aa}, which we only expect to be significant under coherent excitation resonant with the dark exciton energy. $P(x,t)$ represents the incident pumping intensity distribution. This includes Gaussian spots for the continuous wave source and pulsed gate, which also has a Gaussian time dependence. 

The final term in Eq.~(\ref{eq:GP}) accounts for a phenomenological energy relaxation of condensed polaritons: 
\begin{equation} \mathfrak{R}[\psi(x,t)]=-\left(\nu+\nu^\prime|\psi(x,t)|^2\right)\left(\hat{E}_\mathrm{LP}-\mu(x,t)\right)\psi(x,t),\label{eq:relax} 
\end{equation} 

where $\nu$ and $\nu^\prime$ are parameters determining the strength of spontaneous and stimulated energy relaxation, respectively. The energy relaxation has been introduced in this form in several recent works~\cite{Read2009prb,Wouters2010prb,Wouters2012njp,Wertz12prl} with justification arising from consideration of Boltzmann scattering rates \cite{Solnyshkov2013arXiv}. Equivalently, the relaxation is due to the scattering of particles into and out of the condensate, which introduces an imaginary component to the kinetic energy operator as recently demonstrated within a Keldysh functional integral approach \cite{Sieberer2013arxive}. The local effective chemical potential, $\mu(x,t)$, is chosen to enforce particle number conservation (see, e.g., Ref.~\onlinecite{Anton:2013aa} for more details).

For the calculations we used the following parameters: $m=7.3\times10^{-5}\: m_{0}$, with $m_{0}$ the free electron mass, $\alpha=2.4\times10^{-3}$~meV~$\mu$m$^{2}$, $\Gamma=1/18$~ps$^{-1}$, $\Gamma{}_{A}$=1/100~ps$^{-1}$, $\Gamma_{I}=\Gamma_{D}=1/250$~ps$^{-1}$, $t_{R}=t_{D}=0.02$~ps$^{-1}$, $g=0.018$~ps$^{-1}$~$\mu$m$^{2}$, $\hbar r=0.02$~meV~$\mu$m$^{2}$, $\hbar\nu=0.014$, $ $ $\hbar\nu\lyxmathsym{\textquoteright}=0.075$~$\mu$m$^{2}$.

Figure~\ref{fig:TheoryEx} shows the variation with time of the energy spectrum in space. Before the arrival of the gate pulse, polaritons are free to relax down to the collector state. As in the experiment, the application of the gate introduces a potential barrier that blocks polariton propagation. This is temporary, as at later times the excitons in the reservoir at the gate position decay and polaritons are able to once again flow to the collector region. 

For simplicity, the theoretical model assumes perfect Gaussian shaped laser spots and neglects the influence of disorder. Differences in the exact pump profiles and a complicated disorder potential give rise to a slightly different shape of the effective polariton potential in the experiments. However, the overall form of the potential is similar, allowing the theory and experiment to demonstrate the same phenomenology. The presence of disorder in the experiments also gives rise to inhomogeneous broadening, not accounted for in the theory. Fluctuations in the laser intensities may also contribute to broader experimental frequency distributions than in the theory.

\begin{figure}[t]
\setlength{\abovecaptionskip}{-5pt} \setlength{\belowcaptionskip}{-2pt} 

\begin{centering}
\includegraphics[width=1\linewidth]{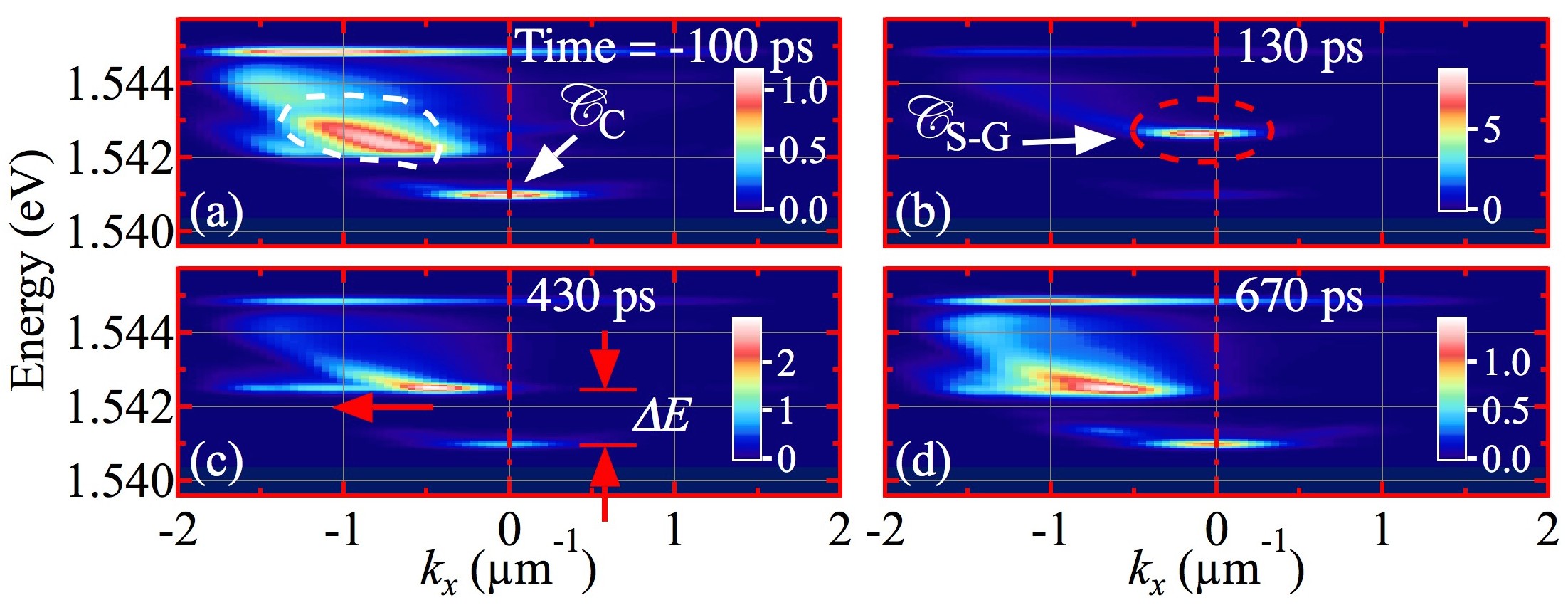} 
\par\end{centering}

\caption{(Color online) Calculated emission intensity false-color scale maps vs. energy and momentum space ($k_{x}$) for different times shown by the labels. $\mathscr{C}_{S-G}$ and $\mathscr{C}_{C}$ mark the trapped condensates between \emph{S-G} and the trapped one at \emph{C} positions in real space, respectively. The acceleration of flowing polaritons is indicated by horizontal red arrows in panel (c). The relative blueshift $\Delta E$ between $\mathscr{C}_{S-G}$ and $\mathscr{C}_{C}$ is indicated by vertical red arrows in panel (c). }

\label{fig:TheoryEk} 
\end{figure}

Figure~\ref{fig:TheoryEk} shows the corresponding energy spectrum in reciprocal space for the same selected times as those used in Fig. \ref{fig:TheoryEx}. The results of the simulations are very similar to the experimental ones shown in Fig.~\ref{fig:spaceK}. Before the arrival of the gate, one can identify polaritons propagating to the left, with negative in-plane wave vector, which have relaxed from the source. At the collector, these polaritons have lost their kinetic energy. The application of the gate blocks the polaritons, such that they remain trapped in a higher-energy state without kinetic energy {[}see Fig.~\ref{fig:TheoryEk}(b){]}. Polaritons become accelerated towards $C$ again as the gate barrier disappears, recovering the initial situation {[}see Fig.~\ref{fig:TheoryEk}(d){]}, in a similar fashion to that observed in the experiments. 

\begin{figure}
\setlength{\abovecaptionskip}{-5pt} \setlength{\belowcaptionskip}{-2pt} 

\begin{centering}
\includegraphics[width=1\linewidth]{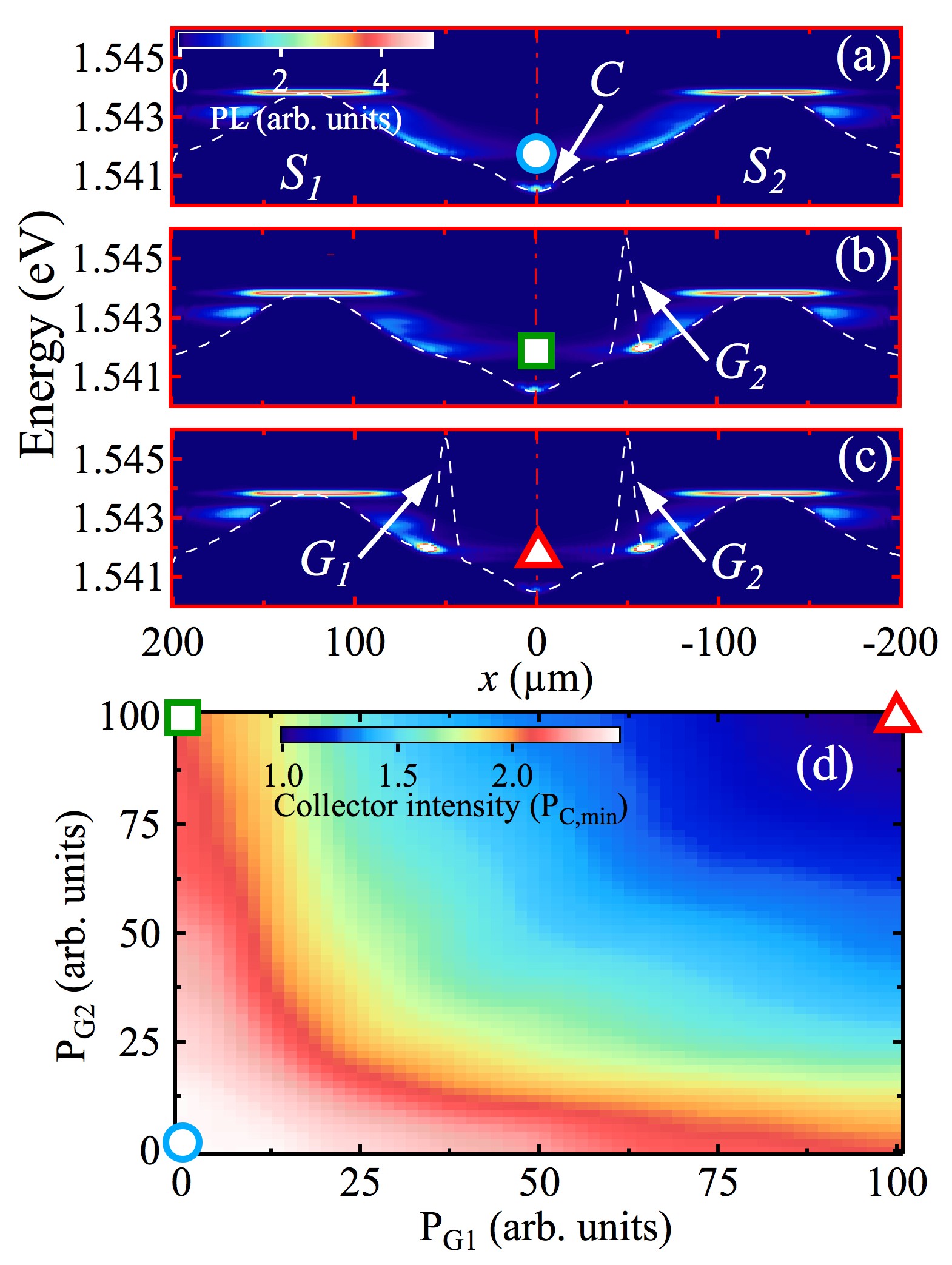} 
\par\end{centering}

\caption{(Color online) (a)-(c) Calculated PL false-color scale maps vs. energy and real space ($x$), for a neural-type logic gate with two inputs, for different gate power combinations: $P_{G1}=P_{G2}>0$, $P_{G1}=P_{G2}=0$ and $P_{G1}>P_{G2}=0$, respectively. (d) Calculated PL at the collector in a false-color scale versus gate powers $P_{G1}$ and $P_{G2}$. The powers used in panels (a)-(c) are marked by a circle, square, and triangle, respectively. }

\label{fig:neural} 
\end{figure}

\subsection{Predictions for multiple switches }

While we have focused on the behavior of an individual condensate transistor switch, future studies will likely be devoted to the linking of multiple elements in analogy to transistors based on coherent excitation \cite{Ballarini13natcomm}. Here, the flexibility in controlling the strength of the individual gate powers may be particularly useful in going beyond the optical replication of CMOS-type logic and considering a neural-type logic inspired by biological networks. For artificial neural networks \cite{rojas96book_neural}, one typically aims to combine the signals of different transistors but with arbitrary controllable weights, which can be engineered to give different functionalities. The sum of the weighted signals is then compared to some threshold to determine the result. This ability is known to allow the implementation of certain tasks with significantly fewer elements than with using chains of typical Boolean logic gates, which is why neural networks can be particularly efficient even if their individual elements may be slow.

As a theoretical example, we consider here the combination of the signals from two condensate transistors in a single polariton channel. Polaritons are excited at two source positions and travel toward the collector region, which is at the mid-point between the two sources. The weighting of the number of polaritons arriving from each source can be independently controlled by two gates. The weighted input from each source is then summed at the collector. In Fig.~\ref{fig:neural}(d), we present the dependence of the collector intensity on the two gate intensities. It is well known that neural-type logic \cite{rojas96book_neural} can reproduce many Boolean-type logic gates. Among the conceivable regimes of operation it is, for example, possible that the device acts as an AND- or OR-type logic gate, depending on the choice of collector threshold (if the collector intensity exceeds some value the result is considered~\textquotedblleft{}1,\textquotedblright{} while if not it is considered~\textquotedblleft{}0\textquotedblright{}). Situations where the collector signal is weak and strong are shown in Figs.~\ref{fig:neural}(a)-\ref{fig:neural}\textbf{(}c), which correspond to the different points marked in Fig.~\ref{fig:neural}\textbf{(}d) by a circle, square, and triangle, respectively. We stress that the combination of two weighted inputs at the collector is only a first step. In principle, the combination of larger numbers of inputs could also be imagined, by designing patterned microcavities where multiple ridges join together.

\section{Conclusions}

\label{sec:conclusions}

In summary, we present a time-resolved PL study in real and momentum space of a polariton switch consisting of a 20 micron-wide ridge. A polariton flow in real space from $S$ to $C$ is gated by a potential barrier induced by optically generated excitons at $G$. By choosing comparatively low gate powers interruption of the flow can be achieved within tens of picoseconds, while maintaining a reasonably high \textit{on-off} signal contrast. However, the inverse process, i.e. switching from the \textit{off} to the \textit{on} state, takes hundreds of picoseconds due to the long-lived excitons at $G$. Numerical simulations based on the modified Gross-Pitaevskii equation reproduce well the dynamics of our device.

\section{Acknowledgements}

\label{sec:acknowledgements}

C.A. acknowledges financial support from Spanish FPU scholarships. P.S. acknowledges Greek GSRT program {}``ARISTEIA\textquotedbl{} (1978) and EU ERC {}``Polaflow'' for financial support. The work was partially supported by the Spanish MEC MAT2011-22997, CAM (S-2009/ESP-1503) and INDEX (289968) projects.


%


\begin{thebibliography}{56}%
\makeatletter
\providecommand \@ifxundefined [1]{%
 \@ifx{#1\undefined}
}%
\providecommand \@ifnum [1]{%
 \ifnum #1\expandafter \@firstoftwo
 \else \expandafter \@secondoftwo
 \fi
}%
\providecommand \@ifx [1]{%
 \ifx #1\expandafter \@firstoftwo
 \else \expandafter \@secondoftwo
 \fi
}%
\providecommand \natexlab [1]{#1}%
\providecommand \enquote  [1]{``#1''}%
\providecommand \bibnamefont  [1]{#1}%
\providecommand \bibfnamefont [1]{#1}%
\providecommand \citenamefont [1]{#1}%
\providecommand \href@noop [0]{\@secondoftwo}%
\providecommand \href [0]{\begingroup \@sanitize@url \@href}%
\providecommand \@href[1]{\@@startlink{#1}\@@href}%
\providecommand \@@href[1]{\endgroup#1\@@endlink}%
\providecommand \@sanitize@url [0]{\catcode `\\12\catcode `\$12\catcode
  `\&12\catcode `\#12\catcode `\^12\catcode `\_12\catcode `\%12\relax}%
\providecommand \@@startlink[1]{}%
\providecommand \@@endlink[0]{}%
\providecommand \url  [0]{\begingroup\@sanitize@url \@url }%
\providecommand \@url [1]{\endgroup\@href {#1}{\urlprefix }}%
\providecommand \urlprefix  [0]{URL }%
\providecommand \Eprint [0]{\href }%
\providecommand \doibase [0]{http://dx.doi.org/}%
\providecommand \selectlanguage [0]{\@gobble}%
\providecommand \bibinfo  [0]{\@secondoftwo}%
\providecommand \bibfield  [0]{\@secondoftwo}%
\providecommand \translation [1]{[#1]}%
\providecommand \BibitemOpen [0]{}%
\providecommand \bibitemStop [0]{}%
\providecommand \bibitemNoStop [0]{.\EOS\space}%
\providecommand \EOS [0]{\spacefactor3000\relax}%
\providecommand \BibitemShut  [1]{\csname bibitem#1\endcsname}%
\let\auto@bib@innerbib\@empty
\bibitem [{\citenamefont {Weisbuch}\ \emph {et~al.}(1992)\citenamefont
  {Weisbuch}, \citenamefont {Nishioka}, \citenamefont {Ishikawa},\ and\
  \citenamefont {Arakawa}}]{weisbuch_observation_1992}%
  \BibitemOpen
  \bibfield  {author} {\bibinfo {author} {\bibfnamefont {C.}~\bibnamefont
  {Weisbuch}}, \bibinfo {author} {\bibfnamefont {M.}~\bibnamefont {Nishioka}},
  \bibinfo {author} {\bibfnamefont {A.}~\bibnamefont {Ishikawa}}, \ and\
  \bibinfo {author} {\bibfnamefont {Y.}~\bibnamefont {Arakawa}},\ }\href
  {\doibase 10.1103/PhysRevLett.69.3314} {\bibfield  {journal} {\bibinfo
  {journal} {Phys. Rev. Lett.}\ }\textbf {\bibinfo {volume} {69}},\ \bibinfo
  {pages} {3314} (\bibinfo {year} {1992})}\BibitemShut {NoStop}%
\bibitem [{\citenamefont {Kasprzak}\ \emph {et~al.}(2006)\citenamefont
  {Kasprzak}, \citenamefont {Richard}, \citenamefont {Kundermann},
  \citenamefont {Baas}, \citenamefont {Jeambrun}, \citenamefont {Keeling},
  \citenamefont {Marchetti}, \citenamefont {Szymanska}, \citenamefont {Andre},
  \citenamefont {Staehli}, \citenamefont {Savona}, \citenamefont {Littlewood},
  \citenamefont {Deveaud},\ and\ \citenamefont {Dang}}]{kasprzak06nat}%
  \BibitemOpen
  \bibfield  {author} {\bibinfo {author} {\bibfnamefont {J.}~\bibnamefont
  {Kasprzak}}, \bibinfo {author} {\bibfnamefont {M.}~\bibnamefont {Richard}},
  \bibinfo {author} {\bibfnamefont {S.}~\bibnamefont {Kundermann}}, \bibinfo
  {author} {\bibfnamefont {A.}~\bibnamefont {Baas}}, \bibinfo {author}
  {\bibfnamefont {P.}~\bibnamefont {Jeambrun}}, \bibinfo {author}
  {\bibfnamefont {J.~M.~J.}\ \bibnamefont {Keeling}}, \bibinfo {author}
  {\bibfnamefont {F.~M.}\ \bibnamefont {Marchetti}}, \bibinfo {author}
  {\bibfnamefont {M.~H.}\ \bibnamefont {Szymanska}}, \bibinfo {author}
  {\bibfnamefont {R.}~\bibnamefont {Andre}}, \bibinfo {author} {\bibfnamefont
  {J.~L.}\ \bibnamefont {Staehli}}, \bibinfo {author} {\bibfnamefont
  {V.}~\bibnamefont {Savona}}, \bibinfo {author} {\bibfnamefont {P.~B.}\
  \bibnamefont {Littlewood}}, \bibinfo {author} {\bibfnamefont
  {B.}~\bibnamefont {Deveaud}}, \ and\ \bibinfo {author} {\bibfnamefont
  {L.~S.}\ \bibnamefont {Dang}},\ }\href {\doibase 10.1038/nature05131}
  {\bibfield  {journal} {\bibinfo  {journal} {Nature}\ }\textbf {\bibinfo
  {volume} {443}},\ \bibinfo {pages} {409} (\bibinfo {year}
  {2006})}\BibitemShut {NoStop}%
\bibitem [{\citenamefont {Cerna}\ \emph {et~al.}(2013)\citenamefont {Cerna},
  \citenamefont {Leger}, \citenamefont {Para{\"\i}so}, \citenamefont {Wouters},
  \citenamefont {Morier-Genoud}, \citenamefont {Portella-Oberli},\ and\
  \citenamefont {Deveaud}}]{Cerna13ncomm}%
  \BibitemOpen
  \bibfield  {author} {\bibinfo {author} {\bibfnamefont {R.}~\bibnamefont
  {Cerna}}, \bibinfo {author} {\bibfnamefont {Y.}~\bibnamefont {Leger}},
  \bibinfo {author} {\bibfnamefont {T.~K.}\ \bibnamefont {Para{\"\i}so}},
  \bibinfo {author} {\bibfnamefont {M.}~\bibnamefont {Wouters}}, \bibinfo
  {author} {\bibfnamefont {F.}~\bibnamefont {Morier-Genoud}}, \bibinfo {author}
  {\bibfnamefont {M.~T.}\ \bibnamefont {Portella-Oberli}}, \ and\ \bibinfo
  {author} {\bibfnamefont {B.}~\bibnamefont {Deveaud}},\ }\href
  {http://dx.doi.org/10.1038/ncomms3008} {\bibfield  {journal} {\bibinfo
  {journal} {Nat. Commun.}\ }\textbf {\bibinfo {volume} {4}},\ \bibinfo {pages}
  {2008} (\bibinfo {year}
  {2013})}\BibitemShut {NoStop}%
\bibitem [{\citenamefont {De~Giorgi}\ \emph {et~al.}(2012)\citenamefont
  {De~Giorgi}, \citenamefont {Ballarini}, \citenamefont {Cancellieri},
  \citenamefont {Marchetti}, \citenamefont {Szymanska}, \citenamefont
  {Tejedor}, \citenamefont {Cingolani}, \citenamefont {Giacobino},
  \citenamefont {Bramati}, \citenamefont {Gigli},\ and\ \citenamefont
  {Sanvitto}}]{De-Giorgi:2012aa}%
  \BibitemOpen
  \bibfield  {author} {\bibinfo {author} {\bibfnamefont {M.}~\bibnamefont
  {De~Giorgi}}, \bibinfo {author} {\bibfnamefont {D.}~\bibnamefont
  {Ballarini}}, \bibinfo {author} {\bibfnamefont {E.}~\bibnamefont
  {Cancellieri}}, \bibinfo {author} {\bibfnamefont {F.~M.}\ \bibnamefont
  {Marchetti}}, \bibinfo {author} {\bibfnamefont {M.~H.}\ \bibnamefont
  {Szymanska}}, \bibinfo {author} {\bibfnamefont {C.}~\bibnamefont {Tejedor}},
  \bibinfo {author} {\bibfnamefont {R.}~\bibnamefont {Cingolani}}, \bibinfo
  {author} {\bibfnamefont {E.}~\bibnamefont {Giacobino}}, \bibinfo {author}
  {\bibfnamefont {A.}~\bibnamefont {Bramati}}, \bibinfo {author} {\bibfnamefont
  {G.}~\bibnamefont {Gigli}}, \ and\ \bibinfo {author} {\bibfnamefont
  {D.}~\bibnamefont {Sanvitto}},\ }\href
  {http://link.aps.org/doi/10.1103/PhysRevLett.109.266407} {\bibfield
  {journal} {\bibinfo  {journal} {Phys. Rev. Lett.}\ }\textbf {\bibinfo
  {volume} {109}},\ \bibinfo {pages} {266407} (\bibinfo {year}
  {2012})}\BibitemShut {NoStop}%
\bibitem [{\citenamefont {Cancellieri}\ \emph {et~al.}(2014)\citenamefont
  {Cancellieri}, \citenamefont {Hayat}, \citenamefont {Steinberg},
  \citenamefont {Giacobino},\ and\ \citenamefont
  {Bramati}}]{Cancellieri14prl-starkShift}%
  \BibitemOpen
  \bibfield  {author} {\bibinfo {author} {\bibfnamefont {E.}~\bibnamefont
  {Cancellieri}}, \bibinfo {author} {\bibfnamefont {A.}~\bibnamefont {Hayat}},
  \bibinfo {author} {\bibfnamefont {A.}~\bibnamefont {Steinberg}}, \bibinfo
  {author} {\bibfnamefont {E.}~\bibnamefont {Giacobino}}, \ and\ \bibinfo
  {author} {\bibfnamefont {A.}~\bibnamefont {Bramati}},\ }\href {\doibase
  10.1103/PhysRevLett.112.053601} {\bibfield  {journal} {\bibinfo  {journal}
  {Phys. Rev. Lett.}\ }\textbf {\bibinfo {volume} {112}},\ \bibinfo {pages}
  {053601} (\bibinfo {year} {2014})}\BibitemShut {NoStop}%
\bibitem [{\citenamefont {Johne}\ \emph {et~al.}(2010)\citenamefont {Johne},
  \citenamefont {Shelykh}, \citenamefont {Solnyshkov},\ and\ \citenamefont
  {Malpuech}}]{johne10prb}%
  \BibitemOpen
  \bibfield  {author} {\bibinfo {author} {\bibfnamefont {R.}~\bibnamefont
  {Johne}}, \bibinfo {author} {\bibfnamefont {I.~A.}\ \bibnamefont {Shelykh}},
  \bibinfo {author} {\bibfnamefont {D.~D.}\ \bibnamefont {Solnyshkov}}, \ and\
  \bibinfo {author} {\bibfnamefont {G.}~\bibnamefont {Malpuech}},\ }\href
  {\doibase 10.1103/PhysRevB.81.125327} {\bibfield  {journal} {\bibinfo
  {journal} {Phys. Rev. B}\ }\textbf {\bibinfo {volume} {81}},\ \bibinfo
  {pages} {125327} (\bibinfo {year} {2010})}\BibitemShut {NoStop}%
\bibitem [{\citenamefont {Shelykh}\ \emph {et~al.}(2010)\citenamefont
  {Shelykh}, \citenamefont {Johne}, \citenamefont {Solnyshkov},\ and\
  \citenamefont {Malpuech}}]{Shelykh10prb-spin-transistor}%
  \BibitemOpen
  \bibfield  {author} {\bibinfo {author} {\bibfnamefont {I.~A.}\ \bibnamefont
  {Shelykh}}, \bibinfo {author} {\bibfnamefont {R.}~\bibnamefont {Johne}},
  \bibinfo {author} {\bibfnamefont {D.~D.}\ \bibnamefont {Solnyshkov}}, \ and\
  \bibinfo {author} {\bibfnamefont {G.}~\bibnamefont {Malpuech}},\ }\href
  {\doibase 10.1103/PhysRevB.82.153303} {\bibfield  {journal} {\bibinfo
  {journal} {Phys. Rev. B}\ }\textbf {\bibinfo {volume} {82}},\ \bibinfo
  {pages} {153303} (\bibinfo {year} {2010})}\BibitemShut {NoStop}%
\bibitem [{\citenamefont {Espinosa-Ortega}\ \emph {et~al.}(2013)\citenamefont
  {Espinosa-Ortega}, \citenamefont {Liew},\ and\ \citenamefont
  {Shelykh}}]{Espinosa-Ortega13apl}%
  \BibitemOpen
  \bibfield  {author} {\bibinfo {author} {\bibfnamefont {T.}~\bibnamefont
  {Espinosa-Ortega}}, \bibinfo {author} {\bibfnamefont {T.~C.~H.}\ \bibnamefont
  {Liew}}, \ and\ \bibinfo {author} {\bibfnamefont {I.~A.}\ \bibnamefont
  {Shelykh}},\ }\href {\doibase http://dx.doi.org/10.1063/1.4829363} {\bibfield
   {journal} {\bibinfo  {journal} {Applied Physics Letters}\ }\textbf {\bibinfo
  {volume} {103}},\ \bibinfo {eid} {191110} (\bibinfo {year}
  {2013})}\BibitemShut {NoStop}%
\bibitem [{\citenamefont {Flayac}\ and\ \citenamefont
  {Savenko}(2013)}]{flayac13apl-polariton-router}%
  \BibitemOpen
  \bibfield  {author} {\bibinfo {author} {\bibfnamefont {H.}~\bibnamefont
  {Flayac}}\ and\ \bibinfo {author} {\bibfnamefont {I.~G.}\ \bibnamefont
  {Savenko}},\ }\href {\doibase http://dx.doi.org/10.1063/1.4830007} {\bibfield
   {journal} {\bibinfo  {journal} {Appl. Phys. Lett.}\ }\textbf {\bibinfo
  {volume} {103}},\ \bibinfo {eid} {201105} (\bibinfo {year}
  {2013})}\BibitemShut {NoStop}%
\bibitem [{\citenamefont {Petrov}\ and\ \citenamefont
  {Kavokin}(2013)}]{petrov13prb}%
  \BibitemOpen
  \bibfield  {author} {\bibinfo {author} {\bibfnamefont {M.~Y.}\ \bibnamefont
  {Petrov}}\ and\ \bibinfo {author} {\bibfnamefont {A.~V.}\ \bibnamefont
  {Kavokin}},\ }\href {\doibase 10.1103/PhysRevB.88.035308} {\bibfield
  {journal} {\bibinfo  {journal} {Phys. Rev. B}\ }\textbf {\bibinfo {volume}
  {88}},\ \bibinfo {pages} {035308} (\bibinfo {year} {2013})}\BibitemShut
  {NoStop}%
\bibitem [{\citenamefont {Liew}\ \emph {et~al.}(2008)\citenamefont {Liew},
  \citenamefont {Kavokin},\ and\ \citenamefont {Shelykh}}]{liew08prl_neurons}%
  \BibitemOpen
  \bibfield  {author} {\bibinfo {author} {\bibfnamefont {T.~C.~H.}\
  \bibnamefont {Liew}}, \bibinfo {author} {\bibfnamefont {A.~V.}\ \bibnamefont
  {Kavokin}}, \ and\ \bibinfo {author} {\bibfnamefont {I.~A.}\ \bibnamefont
  {Shelykh}},\ }\href {http://www.ncbi.nlm.nih.gov/pubmed/18764129} {\bibfield
  {journal} {\bibinfo  {journal} {Phys. Rev. Lett.}\ }\textbf {\bibinfo
  {volume} {101}},\ \bibinfo {pages} {016402} (\bibinfo {year}
  {2008})}\BibitemShut {NoStop}%
\bibitem [{\citenamefont {Espinosa-Ortega}\ and\ \citenamefont
  {Liew}(2013)}]{Espinosa-Ortega:2013aa}%
  \BibitemOpen
  \bibfield  {author} {\bibinfo {author} {\bibfnamefont {T.}~\bibnamefont
  {Espinosa-Ortega}}\ and\ \bibinfo {author} {\bibfnamefont {T.~C.~H.}\
  \bibnamefont {Liew}},\ }\href
  {http://link.aps.org/doi/10.1103/PhysRevB.87.195305} {\bibfield  {journal}
  {\bibinfo  {journal} {Phys. Rev. B}\ }\textbf {\bibinfo {volume} {87}},\
  \bibinfo {pages} {195305} (\bibinfo {year} {2013})}\BibitemShut {NoStop}%
\bibitem [{\citenamefont {Amo}\ \emph {et~al.}(2009)\citenamefont {Amo},
  \citenamefont {Sanvitto}, \citenamefont {Laussy}, \citenamefont {Ballarini},
  \citenamefont {del Valle}, \citenamefont {Martin}, \citenamefont {Lemaitre},
  \citenamefont {Bloch}, \citenamefont {Krizhanovskii}, \citenamefont
  {Skolnick}, \citenamefont {Tejedor},\ and\ \citenamefont {Vina}}]{amo09nat}%
  \BibitemOpen
  \bibfield  {author} {\bibinfo {author} {\bibfnamefont {A.}~\bibnamefont
  {Amo}}, \bibinfo {author} {\bibfnamefont {D.}~\bibnamefont {Sanvitto}},
  \bibinfo {author} {\bibfnamefont {F.~P.}\ \bibnamefont {Laussy}}, \bibinfo
  {author} {\bibfnamefont {D.}~\bibnamefont {Ballarini}}, \bibinfo {author}
  {\bibfnamefont {E.}~\bibnamefont {del Valle}}, \bibinfo {author}
  {\bibfnamefont {M.~D.}\ \bibnamefont {Martin}}, \bibinfo {author}
  {\bibfnamefont {A.}~\bibnamefont {Lemaitre}}, \bibinfo {author}
  {\bibfnamefont {J.}~\bibnamefont {Bloch}}, \bibinfo {author} {\bibfnamefont
  {D.~N.}\ \bibnamefont {Krizhanovskii}}, \bibinfo {author} {\bibfnamefont
  {M.~S.}\ \bibnamefont {Skolnick}}, \bibinfo {author} {\bibfnamefont
  {C.}~\bibnamefont {Tejedor}}, \ and\ \bibinfo {author} {\bibfnamefont
  {L.}~\bibnamefont {Vina}},\ }\href {\doibase 10.1038/nature07640} {\bibfield
  {journal} {\bibinfo  {journal} {Nature}\ }\textbf {\bibinfo {volume} {457}},\
  \bibinfo {pages} {291} (\bibinfo {year} {2009})}\BibitemShut {NoStop}%
\bibitem [{\citenamefont {Adrados}\ \emph {et~al.}(2011)\citenamefont
  {Adrados}, \citenamefont {Liew}, \citenamefont {Amo}, \citenamefont
  {Mart{\'\i}n}, \citenamefont {Sanvitto}, \citenamefont {Ant{\'o}n},
  \citenamefont {Giacobino}, \citenamefont {Kavokin}, \citenamefont {Bramati},\
  and\ \citenamefont {Vi{\~n}a}}]{adrados11prl-bullets}%
  \BibitemOpen
  \bibfield  {author} {\bibinfo {author} {\bibfnamefont {C.}~\bibnamefont
  {Adrados}}, \bibinfo {author} {\bibfnamefont {T.~C.~H.}\ \bibnamefont
  {Liew}}, \bibinfo {author} {\bibfnamefont {A.}~\bibnamefont {Amo}}, \bibinfo
  {author} {\bibfnamefont {M.~D.}\ \bibnamefont {Mart{\'\i}n}}, \bibinfo
  {author} {\bibfnamefont {D.}~\bibnamefont {Sanvitto}}, \bibinfo {author}
  {\bibfnamefont {C.}~\bibnamefont {Ant{\'o}n}}, \bibinfo {author}
  {\bibfnamefont {E.}~\bibnamefont {Giacobino}}, \bibinfo {author}
  {\bibfnamefont {A.}~\bibnamefont {Kavokin}}, \bibinfo {author} {\bibfnamefont
  {A.}~\bibnamefont {Bramati}}, \ and\ \bibinfo {author} {\bibfnamefont
  {L.}~\bibnamefont {Vi{\~n}a}},\ }\href
  {http://link.aps.org/doi/10.1103/PhysRevLett.107.146402} {\bibfield
  {journal} {\bibinfo  {journal} {Phys. Rev. Lett.}\ }\textbf {\bibinfo
  {volume} {107}},\ \bibinfo {pages} {146402} (\bibinfo {year}
  {2011})}\BibitemShut {NoStop}%
\bibitem [{\citenamefont {Amo}\ \emph {et~al.}(2010{\natexlab{a}})\citenamefont
  {Amo}, \citenamefont {Pigeon}, \citenamefont {Adrados}, \citenamefont
  {Houdre}, \citenamefont {Giacobino}, \citenamefont {Ciuti},\ and\
  \citenamefont {Bramati}}]{amo10prb}%
  \BibitemOpen
  \bibfield  {author} {\bibinfo {author} {\bibfnamefont {A.}~\bibnamefont
  {Amo}}, \bibinfo {author} {\bibfnamefont {S.}~\bibnamefont {Pigeon}},
  \bibinfo {author} {\bibfnamefont {C.}~\bibnamefont {Adrados}}, \bibinfo
  {author} {\bibfnamefont {R.}~\bibnamefont {Houdre}}, \bibinfo {author}
  {\bibfnamefont {E.}~\bibnamefont {Giacobino}}, \bibinfo {author}
  {\bibfnamefont {C.}~\bibnamefont {Ciuti}}, \ and\ \bibinfo {author}
  {\bibfnamefont {A.}~\bibnamefont {Bramati}},\ }\href {\doibase
  10.1103/PhysRevB.82.081301} {\bibfield  {journal} {\bibinfo  {journal} {Phys.
  Rev. B}\ }\textbf {\bibinfo {volume} {82}},\ \bibinfo {pages} {081301}
  (\bibinfo {year} {2010}{\natexlab{a}})}\BibitemShut {NoStop}%
\bibitem [{\citenamefont {Martin}\ \emph {et~al.}(2002)\citenamefont {Martin},
  \citenamefont {Aichmayr}, \citenamefont {Vina},\ and\ \citenamefont
  {Andre}}]{martin02prl-polarization-control}%
  \BibitemOpen
  \bibfield  {author} {\bibinfo {author} {\bibfnamefont {M.~D.}\ \bibnamefont
  {Martin}}, \bibinfo {author} {\bibfnamefont {G.}~\bibnamefont {Aichmayr}},
  \bibinfo {author} {\bibfnamefont {L.}~\bibnamefont {Vina}}, \ and\ \bibinfo
  {author} {\bibfnamefont {R.}~\bibnamefont {Andre}},\ }\href {\doibase
  10.1103/PhysRevLett.89.077402} {\bibfield  {journal} {\bibinfo  {journal}
  {Phys. Rev. Lett.}\ }\textbf {\bibinfo {volume} {89}},\ \bibinfo {pages}
  {077402} (\bibinfo {year} {2002})}\BibitemShut {NoStop}%
\bibitem [{\citenamefont {Amo}\ \emph {et~al.}(2010{\natexlab{b}})\citenamefont
  {Amo}, \citenamefont {Liew}, \citenamefont {Adrados}, \citenamefont {Houdre},
  \citenamefont {Giacobino}, \citenamefont {Kavokin},\ and\ \citenamefont
  {Bramati}}]{amo10nphot-spin-switch}%
  \BibitemOpen
  \bibfield  {author} {\bibinfo {author} {\bibfnamefont {A.}~\bibnamefont
  {Amo}}, \bibinfo {author} {\bibfnamefont {T.~C.~H.}\ \bibnamefont {Liew}},
  \bibinfo {author} {\bibfnamefont {C.}~\bibnamefont {Adrados}}, \bibinfo
  {author} {\bibfnamefont {R.}~\bibnamefont {Houdre}}, \bibinfo {author}
  {\bibfnamefont {E.}~\bibnamefont {Giacobino}}, \bibinfo {author}
  {\bibfnamefont {A.~V.}\ \bibnamefont {Kavokin}}, \ and\ \bibinfo {author}
  {\bibfnamefont {A.}~\bibnamefont {Bramati}},\ }\href
  {http://dx.doi.org/10.1038/nphoton.2010.79} {\bibfield  {journal} {\bibinfo
  {journal} {Nat. Photon.}\ }\textbf {\bibinfo {volume} {4}},\ \bibinfo {pages}
  {361} (\bibinfo {year} {2010}{\natexlab{b}})}\BibitemShut {NoStop}%
\bibitem [{\citenamefont {Malpuech}\ \emph {et~al.}(2006)\citenamefont
  {Malpuech}, \citenamefont {Glazov}, \citenamefont {Shelykh}, \citenamefont
  {Bigenwald},\ and\ \citenamefont {Kavokin}}]{malpuech06apl-polarization}%
  \BibitemOpen
  \bibfield  {author} {\bibinfo {author} {\bibfnamefont {G.}~\bibnamefont
  {Malpuech}}, \bibinfo {author} {\bibfnamefont {M.~M.}\ \bibnamefont
  {Glazov}}, \bibinfo {author} {\bibfnamefont {I.~A.}\ \bibnamefont {Shelykh}},
  \bibinfo {author} {\bibfnamefont {P.}~\bibnamefont {Bigenwald}}, \ and\
  \bibinfo {author} {\bibfnamefont {K.~V.}\ \bibnamefont {Kavokin}},\ }\href
  {\doibase http://dx.doi.org/10.1063/1.2183811} {\bibfield  {journal}
  {\bibinfo  {journal} {Appl. Phys. Lett.}\ }\textbf {\bibinfo {volume} {88}},\
  \bibinfo {eid} {111118} (\bibinfo {year} {2006})}\BibitemShut {NoStop}%
\bibitem [{\citenamefont {{Grosso}}\ \emph {et~al.}()\citenamefont {{Grosso}},
  \citenamefont {{Trebaol}}, \citenamefont {{Wouters}}, \citenamefont
  {{Morier-Genoud}}, \citenamefont {{Portella-Oberli}},\ and\ \citenamefont
  {{Deveaud}}}]{grosso13arXiv-polariton-switch}%
  \BibitemOpen
  \bibfield  {author} {\bibinfo {author} {\bibfnamefont {G.}~\bibnamefont
  {{Grosso}}}, \bibinfo {author} {\bibfnamefont {S.}~\bibnamefont {{Trebaol}}},
  \bibinfo {author} {\bibfnamefont {M.}~\bibnamefont {{Wouters}}}, \bibinfo
  {author} {\bibfnamefont {F.}~\bibnamefont {{Morier-Genoud}}}, \bibinfo
  {author} {\bibfnamefont {M.~T.}\ \bibnamefont {{Portella-Oberli}}}, \ and\
  \bibinfo {author} {\bibfnamefont {B.}~\bibnamefont {{Deveaud}}},\ }\href@noop
  {} {\ }\Eprint {http://arxiv.org/abs/1312.1238} {arXiv:1312.1238
  [cond-mat.quant-gas]} \BibitemShut {NoStop}%
\bibitem [{\citenamefont {Para{\"\i}so}\ \emph {et~al.}(2010)\citenamefont
  {Para{\"\i}so}, \citenamefont {Wouters}, \citenamefont {L{\'e}ger},
  \citenamefont {Morier-Genoud},\ and\ \citenamefont
  {Deveaud-Pl{\'e}dran}}]{Paraiso:2010aa}%
  \BibitemOpen
  \bibfield  {author} {\bibinfo {author} {\bibfnamefont {T.~K.}\ \bibnamefont
  {Para{\"\i}so}}, \bibinfo {author} {\bibfnamefont {M.}~\bibnamefont
  {Wouters}}, \bibinfo {author} {\bibfnamefont {Y.}~\bibnamefont {L{\'e}ger}},
  \bibinfo {author} {\bibfnamefont {F.}~\bibnamefont {Morier-Genoud}}, \ and\
  \bibinfo {author} {\bibfnamefont {B.}~\bibnamefont {Deveaud-Pl{\'e}dran}},\
  }\href {http://dx.doi.org/10.1038/nmat2787} {\bibfield  {journal} {\bibinfo
  {journal} {Nat. Mater.}\ }\textbf {\bibinfo {volume} {9}},\ \bibinfo {pages}
  {655} (\bibinfo {year} {2010})}\BibitemShut {NoStop}%
\bibitem [{\citenamefont {Gippius}\ \emph {et~al.}(2007)\citenamefont
  {Gippius}, \citenamefont {Shelykh}, \citenamefont {Solnyshkov}, \citenamefont
  {Gavrilov}, \citenamefont {Rubo}, \citenamefont {Kavokin}, \citenamefont
  {Tikhodeev},\ and\ \citenamefont {Malpuech}}]{gippius07prl-multistability}%
  \BibitemOpen
  \bibfield  {author} {\bibinfo {author} {\bibfnamefont {N.~A.}\ \bibnamefont
  {Gippius}}, \bibinfo {author} {\bibfnamefont {I.~A.}\ \bibnamefont
  {Shelykh}}, \bibinfo {author} {\bibfnamefont {D.~D.}\ \bibnamefont
  {Solnyshkov}}, \bibinfo {author} {\bibfnamefont {S.~S.}\ \bibnamefont
  {Gavrilov}}, \bibinfo {author} {\bibfnamefont {Y.~G.}\ \bibnamefont {Rubo}},
  \bibinfo {author} {\bibfnamefont {A.~V.}\ \bibnamefont {Kavokin}}, \bibinfo
  {author} {\bibfnamefont {S.~G.}\ \bibnamefont {Tikhodeev}}, \ and\ \bibinfo
  {author} {\bibfnamefont {G.}~\bibnamefont {Malpuech}},\ }\href {\doibase
  10.1103/PhysRevLett.98.236401} {\bibfield  {journal} {\bibinfo  {journal}
  {Phys. Rev. Lett.}\ }\textbf {\bibinfo {volume} {98}},\ \bibinfo {pages}
  {236401} (\bibinfo {year} {2007})}\BibitemShut {NoStop}%
\bibitem [{\citenamefont {Gavrilov}\ \emph {et~al.}(2013)\citenamefont
  {Gavrilov}, \citenamefont {Sekretenko}, \citenamefont {Novikov},
  \citenamefont {Schneider}, \citenamefont {H{\"o}fling}, \citenamefont {Kamp},
  \citenamefont {Forchel},\ and\ \citenamefont
  {Kulakovskii}}]{gavrilov13apl-multistability}%
  \BibitemOpen
  \bibfield  {author} {\bibinfo {author} {\bibfnamefont {S.~S.}\ \bibnamefont
  {Gavrilov}}, \bibinfo {author} {\bibfnamefont {A.~V.}\ \bibnamefont
  {Sekretenko}}, \bibinfo {author} {\bibfnamefont {S.~I.}\ \bibnamefont
  {Novikov}}, \bibinfo {author} {\bibfnamefont {C.}~\bibnamefont {Schneider}},
  \bibinfo {author} {\bibfnamefont {S.}~\bibnamefont {H{\"o}fling}}, \bibinfo
  {author} {\bibfnamefont {M.}~\bibnamefont {Kamp}}, \bibinfo {author}
  {\bibfnamefont {A.}~\bibnamefont {Forchel}}, \ and\ \bibinfo {author}
  {\bibfnamefont {V.~D.}\ \bibnamefont {Kulakovskii}},\ }\href {\doibase
  http://dx.doi.org/10.1063/1.4773523} {\bibfield  {journal} {\bibinfo
  {journal} {Appl. Phys. Lett.}\ }\textbf {\bibinfo {volume} {102}},\ \bibinfo
  {eid} {011104} (\bibinfo {year} {2013})}\BibitemShut {NoStop}%
\bibitem [{\citenamefont {Wertz}\ \emph {et~al.}(2010)\citenamefont {Wertz},
  \citenamefont {Ferrier}, \citenamefont {Solnyshkov}, \citenamefont {Johne},
  \citenamefont {Sanvitto}, \citenamefont {Lemaitre}, \citenamefont {Sagnes},
  \citenamefont {Grousson}, \citenamefont {Kavokin}, \citenamefont {Senellart},
  \citenamefont {Malpuech},\ and\ \citenamefont {Bloch}}]{wertz10nphys-1D}%
  \BibitemOpen
  \bibfield  {author} {\bibinfo {author} {\bibfnamefont {E.}~\bibnamefont
  {Wertz}}, \bibinfo {author} {\bibfnamefont {L.}~\bibnamefont {Ferrier}},
  \bibinfo {author} {\bibfnamefont {D.~D.}\ \bibnamefont {Solnyshkov}},
  \bibinfo {author} {\bibfnamefont {R.}~\bibnamefont {Johne}}, \bibinfo
  {author} {\bibfnamefont {D.}~\bibnamefont {Sanvitto}}, \bibinfo {author}
  {\bibfnamefont {A.}~\bibnamefont {Lemaitre}}, \bibinfo {author}
  {\bibfnamefont {I.}~\bibnamefont {Sagnes}}, \bibinfo {author} {\bibfnamefont
  {R.}~\bibnamefont {Grousson}}, \bibinfo {author} {\bibfnamefont {A.~V.}\
  \bibnamefont {Kavokin}}, \bibinfo {author} {\bibfnamefont {P.}~\bibnamefont
  {Senellart}}, \bibinfo {author} {\bibfnamefont {G.}~\bibnamefont {Malpuech}},
  \ and\ \bibinfo {author} {\bibfnamefont {J.}~\bibnamefont {Bloch}},\ }\href
  {http://dx.doi.org/10.1038/nphys1750} {\bibfield  {journal} {\bibinfo
  {journal} {Nat. Phys.}\ }\textbf {\bibinfo {volume} {6}},\ \bibinfo {pages}
  {860} (\bibinfo {year} {2010})}\BibitemShut {NoStop}%
\bibitem [{\citenamefont {Wertz}\ \emph {et~al.}(2012)\citenamefont {Wertz},
  \citenamefont {Amo}, \citenamefont {Solnyshkov}, \citenamefont {Ferrier},
  \citenamefont {Liew}, \citenamefont {Sanvitto}, \citenamefont {Senellart},
  \citenamefont {Sagnes}, \citenamefont {Lemaitre}, \citenamefont {Kavokin},
  \citenamefont {Malpuech},\ and\ \citenamefont {Bloch}}]{Wertz12prl}%
  \BibitemOpen
  \bibfield  {author} {\bibinfo {author} {\bibfnamefont {E.}~\bibnamefont
  {Wertz}}, \bibinfo {author} {\bibfnamefont {A.}~\bibnamefont {Amo}}, \bibinfo
  {author} {\bibfnamefont {D.~D.}\ \bibnamefont {Solnyshkov}}, \bibinfo
  {author} {\bibfnamefont {L.}~\bibnamefont {Ferrier}}, \bibinfo {author}
  {\bibfnamefont {T.~C.~H.}\ \bibnamefont {Liew}}, \bibinfo {author}
  {\bibfnamefont {D.}~\bibnamefont {Sanvitto}}, \bibinfo {author}
  {\bibfnamefont {P.}~\bibnamefont {Senellart}}, \bibinfo {author}
  {\bibfnamefont {I.}~\bibnamefont {Sagnes}}, \bibinfo {author} {\bibfnamefont
  {A.}~\bibnamefont {Lemaitre}}, \bibinfo {author} {\bibfnamefont {A.~V.}\
  \bibnamefont {Kavokin}}, \bibinfo {author} {\bibfnamefont {G.}~\bibnamefont
  {Malpuech}}, \ and\ \bibinfo {author} {\bibfnamefont {J.}~\bibnamefont
  {Bloch}},\ }\href {\doibase 10.1103/PhysRevLett.109.216404} {\bibfield
  {journal} {\bibinfo  {journal} {Phys. Rev. Lett.}\ }\textbf {\bibinfo
  {volume} {109}},\ \bibinfo {pages} {216404} (\bibinfo {year}
  {2012})}\BibitemShut {NoStop}%
\bibitem [{\citenamefont {Nguyen}\ \emph {et~al.}(2013)\citenamefont {Nguyen},
  \citenamefont {Vishnevsky}, \citenamefont {Sturm}, \citenamefont {Tanese},
  \citenamefont {Solnyshkov}, \citenamefont {Galopin}, \citenamefont
  {Lema{\^\i}tre}, \citenamefont {Sagnes}, \citenamefont {Amo}, \citenamefont
  {Malpuech},\ and\ \citenamefont {Bloch}}]{Nguyen:2013aa}%
  \BibitemOpen
  \bibfield  {author} {\bibinfo {author} {\bibfnamefont {H.~S.}\ \bibnamefont
  {Nguyen}}, \bibinfo {author} {\bibfnamefont {D.}~\bibnamefont {Vishnevsky}},
  \bibinfo {author} {\bibfnamefont {C.}~\bibnamefont {Sturm}}, \bibinfo
  {author} {\bibfnamefont {D.}~\bibnamefont {Tanese}}, \bibinfo {author}
  {\bibfnamefont {D.}~\bibnamefont {Solnyshkov}}, \bibinfo {author}
  {\bibfnamefont {E.}~\bibnamefont {Galopin}}, \bibinfo {author} {\bibfnamefont
  {A.}~\bibnamefont {Lema{\^\i}tre}}, \bibinfo {author} {\bibfnamefont
  {I.}~\bibnamefont {Sagnes}}, \bibinfo {author} {\bibfnamefont
  {A.}~\bibnamefont {Amo}}, \bibinfo {author} {\bibfnamefont {G.}~\bibnamefont
  {Malpuech}}, \ and\ \bibinfo {author} {\bibfnamefont {J.}~\bibnamefont
  {Bloch}},\ }\href {http://link.aps.org/doi/10.1103/PhysRevLett.110.236601}
  {\bibfield  {journal} {\bibinfo  {journal} {Phys. Rev. Lett.}\ }\textbf
  {\bibinfo {volume} {110}},\ \bibinfo {pages} {236601} (\bibinfo {year}
  {2013})}\BibitemShut {NoStop}%
\bibitem [{\citenamefont {Sturm}\ \emph {et~al.}(2014)\citenamefont {Sturm},
  \citenamefont {Tanese}, \citenamefont {Nguyen}, \citenamefont {Flayac},
  \citenamefont {Galopin}, \citenamefont {Lema{\^\i}tre}, \citenamefont
  {Sagnes}, \citenamefont {Solnyshkov}, \citenamefont {Amo}, \citenamefont
  {Malpuech},\ and\ \citenamefont {Bloch}}]{Sturm:2014aa}%
  \BibitemOpen
  \bibfield  {author} {\bibinfo {author} {\bibfnamefont {C.}~\bibnamefont
  {Sturm}}, \bibinfo {author} {\bibfnamefont {D.}~\bibnamefont {Tanese}},
  \bibinfo {author} {\bibfnamefont {H.~S.}\ \bibnamefont {Nguyen}}, \bibinfo
  {author} {\bibfnamefont {H.}~\bibnamefont {Flayac}}, \bibinfo {author}
  {\bibfnamefont {E.}~\bibnamefont {Galopin}}, \bibinfo {author} {\bibfnamefont
  {A.}~\bibnamefont {Lema{\^\i}tre}}, \bibinfo {author} {\bibfnamefont
  {I.}~\bibnamefont {Sagnes}}, \bibinfo {author} {\bibfnamefont
  {D.}~\bibnamefont {Solnyshkov}}, \bibinfo {author} {\bibfnamefont
  {A.}~\bibnamefont {Amo}}, \bibinfo {author} {\bibfnamefont {G.}~\bibnamefont
  {Malpuech}}, \ and\ \bibinfo {author} {\bibfnamefont {J.}~\bibnamefont
  {Bloch}},\ }\href {http://dx.doi.org/10.1038/ncomms4278} {\bibfield
  {journal} {\bibinfo  {journal} {Nat. Commun.}\ }\textbf {\bibinfo {volume}
  {5}},\ \bibinfo {pages}
  {1778} (\bibinfo {year} {2014})}\BibitemShut {NoStop}%
\bibitem [{\citenamefont {Gao}\ \emph {et~al.}(2012)\citenamefont {Gao},
  \citenamefont {Eldridge}, \citenamefont {Liew}, \citenamefont {Tsintzos},
  \citenamefont {Stavrinidis}, \citenamefont {Deligeorgis}, \citenamefont
  {Hatzopoulos},\ and\ \citenamefont
  {Savvidis}}]{gao12prb-Polariton-transistor}%
  \BibitemOpen
  \bibfield  {author} {\bibinfo {author} {\bibfnamefont {T.}~\bibnamefont
  {Gao}}, \bibinfo {author} {\bibfnamefont {P.~S.}\ \bibnamefont {Eldridge}},
  \bibinfo {author} {\bibfnamefont {T.~C.~H.}\ \bibnamefont {Liew}}, \bibinfo
  {author} {\bibfnamefont {S.~I.}\ \bibnamefont {Tsintzos}}, \bibinfo {author}
  {\bibfnamefont {G.}~\bibnamefont {Stavrinidis}}, \bibinfo {author}
  {\bibfnamefont {G.}~\bibnamefont {Deligeorgis}}, \bibinfo {author}
  {\bibfnamefont {Z.}~\bibnamefont {Hatzopoulos}}, \ and\ \bibinfo {author}
  {\bibfnamefont {P.~G.}\ \bibnamefont {Savvidis}},\ }\href
  {http://link.aps.org/doi/10.1103/PhysRevB.85.235102} {\bibfield  {journal}
  {\bibinfo  {journal} {Phys. Rev. B}\ }\textbf {\bibinfo {volume} {85}},\
  \bibinfo {pages} {235102} (\bibinfo {year} {2012})}\BibitemShut {NoStop}%
\bibitem [{\citenamefont {Ant{\'o}n}\ \emph {et~al.}(2012)\citenamefont
  {Ant{\'o}n}, \citenamefont {Liew}, \citenamefont {Tosi}, \citenamefont
  {Mart{\'\i}n}, \citenamefont {Gao}, \citenamefont {Hatzopoulos},
  \citenamefont {Eldridge}, \citenamefont {Savvidis},\ and\ \citenamefont
  {Vi{\~n}a}}]{Anton:2012aa}%
  \BibitemOpen
  \bibfield  {author} {\bibinfo {author} {\bibfnamefont {C.}~\bibnamefont
  {Ant{\'o}n}}, \bibinfo {author} {\bibfnamefont {T.~C.~H.}\ \bibnamefont
  {Liew}}, \bibinfo {author} {\bibfnamefont {G.}~\bibnamefont {Tosi}}, \bibinfo
  {author} {\bibfnamefont {M.~D.}\ \bibnamefont {Mart{\'\i}n}}, \bibinfo
  {author} {\bibfnamefont {T.}~\bibnamefont {Gao}}, \bibinfo {author}
  {\bibfnamefont {Z.}~\bibnamefont {Hatzopoulos}}, \bibinfo {author}
  {\bibfnamefont {P.~S.}\ \bibnamefont {Eldridge}}, \bibinfo {author}
  {\bibfnamefont {P.~G.}\ \bibnamefont {Savvidis}}, \ and\ \bibinfo {author}
  {\bibfnamefont {L.}~\bibnamefont {Vi{\~n}a}},\ }\href
  {http://scitation.aip.org/content/aip/journal/apl/101/26/10.1063/1.4773376}
  {\bibfield  {journal} {\bibinfo  {journal} {Appl. Phys. Lett.}\ }\textbf
  {\bibinfo {volume} {101}},\ \bibinfo {pages}
  {261116} (\bibinfo {year} {2012})}\BibitemShut {NoStop}%
\bibitem [{\citenamefont {Ant{\'o}n}\ \emph
  {et~al.}(2013{\natexlab{a}})\citenamefont {Ant{\'o}n}, \citenamefont {Liew},
  \citenamefont {Tosi}, \citenamefont {Mart{\'\i}n}, \citenamefont {Gao},
  \citenamefont {Hatzopoulos}, \citenamefont {Eldridge}, \citenamefont
  {Savvidis},\ and\ \citenamefont {Vi{\~n}a}}]{Anton:2013aa}%
  \BibitemOpen
  \bibfield  {author} {\bibinfo {author} {\bibfnamefont {C.}~\bibnamefont
  {Ant{\'o}n}}, \bibinfo {author} {\bibfnamefont {T.~C.~H.}\ \bibnamefont
  {Liew}}, \bibinfo {author} {\bibfnamefont {G.}~\bibnamefont {Tosi}}, \bibinfo
  {author} {\bibfnamefont {M.~D.}\ \bibnamefont {Mart{\'\i}n}}, \bibinfo
  {author} {\bibfnamefont {T.}~\bibnamefont {Gao}}, \bibinfo {author}
  {\bibfnamefont {Z.}~\bibnamefont {Hatzopoulos}}, \bibinfo {author}
  {\bibfnamefont {P.~S.}\ \bibnamefont {Eldridge}}, \bibinfo {author}
  {\bibfnamefont {P.~G.}\ \bibnamefont {Savvidis}}, \ and\ \bibinfo {author}
  {\bibfnamefont {L.}~\bibnamefont {Vi{\~n}a}},\ }\href
  {http://link.aps.org/doi/10.1103/PhysRevB.88.035313} {\bibfield  {journal}
  {\bibinfo  {journal} {Phys. Rev. B}\ }\textbf {\bibinfo {volume} {88}},\
  \bibinfo {pages} {035313} (\bibinfo {year} {2013}{\natexlab{a}})}\BibitemShut
  {NoStop}%
\bibitem [{\citenamefont {Tsotsis}\ \emph {et~al.}(2012)\citenamefont
  {Tsotsis}, \citenamefont {Eldridge}, \citenamefont {Gao}, \citenamefont
  {Tsintzos}, \citenamefont {Hatzopoulos},\ and\ \citenamefont
  {Savvidis}}]{Tsotsis12njp}%
  \BibitemOpen
  \bibfield  {author} {\bibinfo {author} {\bibfnamefont {P.}~\bibnamefont
  {Tsotsis}}, \bibinfo {author} {\bibfnamefont {P.~S.}\ \bibnamefont
  {Eldridge}}, \bibinfo {author} {\bibfnamefont {T.}~\bibnamefont {Gao}},
  \bibinfo {author} {\bibfnamefont {S.~I.}\ \bibnamefont {Tsintzos}}, \bibinfo
  {author} {\bibfnamefont {Z.}~\bibnamefont {Hatzopoulos}}, \ and\ \bibinfo
  {author} {\bibfnamefont {P.~G.}\ \bibnamefont {Savvidis}},\ }\href
  {http://stacks.iop.org/1367-2630/14/i=2/a=023060} {\bibfield  {journal}
  {\bibinfo  {journal} {New J. Phys.}\ }\textbf {\bibinfo {volume} {14}},\
  \bibinfo {pages} {023060} (\bibinfo {year} {2012})}\BibitemShut {NoStop}%
\bibitem [{\citenamefont {Tartakovskii}\ \emph {et~al.}(1998)\citenamefont
  {Tartakovskii}, \citenamefont {Kulakovskii}, \citenamefont {Forchel},\ and\
  \citenamefont {Reithmaier}}]{tartakovskii98prb-1d}%
  \BibitemOpen
  \bibfield  {author} {\bibinfo {author} {\bibfnamefont {A.~I.}\ \bibnamefont
  {Tartakovskii}}, \bibinfo {author} {\bibfnamefont {V.~D.}\ \bibnamefont
  {Kulakovskii}}, \bibinfo {author} {\bibfnamefont {A.}~\bibnamefont
  {Forchel}}, \ and\ \bibinfo {author} {\bibfnamefont {J.~P.}\ \bibnamefont
  {Reithmaier}},\ }\href {\doibase 10.1103/PhysRevB.57.R6807} {\bibfield
  {journal} {\bibinfo  {journal} {Phys. Rev. B}\ }\textbf {\bibinfo {volume}
  {57}},\ \bibinfo {pages} {R6807} (\bibinfo {year} {1998})}\BibitemShut
  {NoStop}%
\bibitem [{\citenamefont {Richard}\ \emph {et~al.}(2005)\citenamefont
  {Richard}, \citenamefont {Kasprzak}, \citenamefont {Romestain}, \citenamefont
  {Andr{\'e}},\ and\ \citenamefont {Dang}}]{Richard:2005aa}%
  \BibitemOpen
  \bibfield  {author} {\bibinfo {author} {\bibfnamefont {M.}~\bibnamefont
  {Richard}}, \bibinfo {author} {\bibfnamefont {J.}~\bibnamefont {Kasprzak}},
  \bibinfo {author} {\bibfnamefont {R.}~\bibnamefont {Romestain}}, \bibinfo
  {author} {\bibfnamefont {R.}~\bibnamefont {Andr{\'e}}}, \ and\ \bibinfo
  {author} {\bibfnamefont {L.~S.}\ \bibnamefont {Dang}},\ }\href
  {http://link.aps.org/doi/10.1103/PhysRevLett.94.187401} {\bibfield  {journal}
  {\bibinfo  {journal} {Phys. Rev. Lett.}\ }\textbf {\bibinfo {volume} {94}},\
  \bibinfo {pages} {187401} (\bibinfo {year} {2005})}\BibitemShut {NoStop}%
\bibitem [{\citenamefont {Ant{\'o}n}\ \emph
  {et~al.}(2013{\natexlab{b}})\citenamefont {Ant{\'o}n}, \citenamefont {Liew},
  \citenamefont {Cuadra}, \citenamefont {Mart{\'\i}n}, \citenamefont
  {Eldridge}, \citenamefont {Hatzopoulos}, \citenamefont {Stavrinidis},
  \citenamefont {Savvidis},\ and\ \citenamefont {Vi{\~n}a}}]{Anton:2013ab}%
  \BibitemOpen
  \bibfield  {author} {\bibinfo {author} {\bibfnamefont {C.}~\bibnamefont
  {Ant{\'o}n}}, \bibinfo {author} {\bibfnamefont {T.~C.~H.}\ \bibnamefont
  {Liew}}, \bibinfo {author} {\bibfnamefont {J.}~\bibnamefont {Cuadra}},
  \bibinfo {author} {\bibfnamefont {M.~D.}\ \bibnamefont {Mart{\'\i}n}},
  \bibinfo {author} {\bibfnamefont {P.~S.}\ \bibnamefont {Eldridge}}, \bibinfo
  {author} {\bibfnamefont {Z.}~\bibnamefont {Hatzopoulos}}, \bibinfo {author}
  {\bibfnamefont {G.}~\bibnamefont {Stavrinidis}}, \bibinfo {author}
  {\bibfnamefont {P.~G.}\ \bibnamefont {Savvidis}}, \ and\ \bibinfo {author}
  {\bibfnamefont {L.}~\bibnamefont {Vi{\~n}a}},\ }\href
  {http://link.aps.org/doi/10.1103/PhysRevB.88.245307} {\bibfield  {journal}
  {\bibinfo  {journal} {Phys. Rev. B}\ }\textbf {\bibinfo {volume} {88}},\
  \bibinfo {pages} {245307} (\bibinfo {year} {2013}{\natexlab{b}})}\BibitemShut
  {NoStop}%
\bibitem [{Note1()}]{Note1}%
  \BibitemOpen
  \bibinfo {note} {There are also some other factors such as the relative distance between $S-G$ and $S-C$ that are out of the scope of this work.}\BibitemShut {Stop}%
\bibitem [{\citenamefont {Amo}\ \emph {et~al.}(2010{\natexlab{c}})\citenamefont
  {Amo}, \citenamefont {Sanvitto},\ and\ \citenamefont {Vi{\~n}a}}]{amo10sst}%
  \BibitemOpen
  \bibfield  {author} {\bibinfo {author} {\bibfnamefont {A.}~\bibnamefont
  {Amo}}, \bibinfo {author} {\bibfnamefont {D.}~\bibnamefont {Sanvitto}}, \
  and\ \bibinfo {author} {\bibfnamefont {L.}~\bibnamefont {Vi{\~n}a}},\ }\href
  {http://stacks.iop.org/0268-1242/25/i=4/a=043001} {\bibfield  {journal}
  {\bibinfo  {journal} {Semicond. Sci. Technol.}\ }\textbf {\bibinfo {volume}
  {25}},\ \bibinfo {pages} {043001} (\bibinfo {year}
  {2010}{\natexlab{c}})}\BibitemShut {NoStop}%
\bibitem [{\citenamefont {Leyder}\ \emph {et~al.}(2007)\citenamefont {Leyder},
  \citenamefont {Liew}, \citenamefont {Kavokin}, \citenamefont {Shelykh},
  \citenamefont {Romanelli}, \citenamefont {Karr}, \citenamefont {Giacobino},\
  and\ \citenamefont {Bramati}}]{Leyder2007prl}%
  \BibitemOpen
  \bibfield  {author} {\bibinfo {author} {\bibfnamefont {C.}~\bibnamefont
  {Leyder}}, \bibinfo {author} {\bibfnamefont {T.~C.~H.}\ \bibnamefont {Liew}},
  \bibinfo {author} {\bibfnamefont {A.~V.}\ \bibnamefont {Kavokin}}, \bibinfo
  {author} {\bibfnamefont {I.~A.}\ \bibnamefont {Shelykh}}, \bibinfo {author}
  {\bibfnamefont {M.}~\bibnamefont {Romanelli}}, \bibinfo {author}
  {\bibfnamefont {J.~P.}\ \bibnamefont {Karr}}, \bibinfo {author}
  {\bibfnamefont {E.}~\bibnamefont {Giacobino}}, \ and\ \bibinfo {author}
  {\bibfnamefont {A.}~\bibnamefont {Bramati}},\ }\href {\doibase
  10.1103/PhysRevLett.99.196402} {\bibfield  {journal} {\bibinfo  {journal}
  {Phys. Rev. Lett.}\ }\textbf {\bibinfo {volume} {99}},\ \bibinfo {pages}
  {196402} (\bibinfo {year} {2007})}\BibitemShut {NoStop}%
\bibitem [{\citenamefont {Ballarini}\ \emph {et~al.}(2013)\citenamefont
  {Ballarini}, \citenamefont {De~Giorgi}, \citenamefont {Cancellieri},
  \citenamefont {Houdr{\'e}}, \citenamefont {Giacobino}, \citenamefont
  {Cingolani}, \citenamefont {Bramati}, \citenamefont {Gigli},\ and\
  \citenamefont {Sanvitto}}]{Ballarini13natcomm}%
  \BibitemOpen
  \bibfield  {author} {\bibinfo {author} {\bibfnamefont {D.}~\bibnamefont
  {Ballarini}}, \bibinfo {author} {\bibfnamefont {M.}~\bibnamefont
  {De~Giorgi}}, \bibinfo {author} {\bibfnamefont {E.}~\bibnamefont
  {Cancellieri}}, \bibinfo {author} {\bibfnamefont {R.}~\bibnamefont
  {Houdr{\'e}}}, \bibinfo {author} {\bibfnamefont {E.}~\bibnamefont
  {Giacobino}}, \bibinfo {author} {\bibfnamefont {R.}~\bibnamefont
  {Cingolani}}, \bibinfo {author} {\bibfnamefont {A.}~\bibnamefont {Bramati}},
  \bibinfo {author} {\bibfnamefont {G.}~\bibnamefont {Gigli}}, \ and\ \bibinfo
  {author} {\bibfnamefont {D.}~\bibnamefont {Sanvitto}},\ }\href
  {http://dx.doi.org/10.1038/ncomms2734} {\bibfield  {journal} {\bibinfo
  {journal} {Nat. Commun.}\ }\textbf {\bibinfo {volume} {4}},\ \bibinfo {pages}
  {1778} (\bibinfo {year} {2013})}\BibitemShut {NoStop}%
\bibitem [{\citenamefont {Hayat}\ \emph {et~al.}(2012)\citenamefont {Hayat},
  \citenamefont {Lange}, \citenamefont {Rozema}, \citenamefont {Darabi},
  \citenamefont {van Driel}, \citenamefont {Steinberg}, \citenamefont {Nelsen},
  \citenamefont {Snoke}, \citenamefont {Pfeiffer},\ and\ \citenamefont
  {West}}]{Hayat12prl-dynamic-stark-effect}%
  \BibitemOpen
  \bibfield  {author} {\bibinfo {author} {\bibfnamefont {A.}~\bibnamefont
  {Hayat}}, \bibinfo {author} {\bibfnamefont {C.}~\bibnamefont {Lange}},
  \bibinfo {author} {\bibfnamefont {L.~A.}\ \bibnamefont {Rozema}}, \bibinfo
  {author} {\bibfnamefont {A.}~\bibnamefont {Darabi}}, \bibinfo {author}
  {\bibfnamefont {H.~M.}\ \bibnamefont {van Driel}}, \bibinfo {author}
  {\bibfnamefont {A.~M.}\ \bibnamefont {Steinberg}}, \bibinfo {author}
  {\bibfnamefont {B.}~\bibnamefont {Nelsen}}, \bibinfo {author} {\bibfnamefont
  {D.~W.}\ \bibnamefont {Snoke}}, \bibinfo {author} {\bibfnamefont {L.~N.}\
  \bibnamefont {Pfeiffer}}, \ and\ \bibinfo {author} {\bibfnamefont {K.~W.}\
  \bibnamefont {West}},\ }\href {\doibase 10.1103/PhysRevLett.109.033605}
  {\bibfield  {journal} {\bibinfo  {journal} {Phys. Rev. Lett.}\ }\textbf
  {\bibinfo {volume} {109}},\ \bibinfo {pages} {033605} (\bibinfo {year}
  {2012})}\BibitemShut {NoStop}%
\bibitem [{Note2()}]{Note2}%
  \BibitemOpen
  \bibinfo {note} {See Supplemental Material at http://link.aps.org/supplemental/10.1103/PhysRevB.89.235312 for movies showing polariton dynamics corresponding to Figs. 2, 4, and 8.}\BibitemShut {Stop}%
\bibitem [{\citenamefont {Balili}\ \emph {et~al.}(2007)\citenamefont {Balili},
  \citenamefont {Hartwell}, \citenamefont {Snoke}, \citenamefont {Pfeiffer},\
  and\ \citenamefont {West}}]{Balili2007science}%
  \BibitemOpen
  \bibfield  {author} {\bibinfo {author} {\bibfnamefont {R.}~\bibnamefont
  {Balili}}, \bibinfo {author} {\bibfnamefont {V.}~\bibnamefont {Hartwell}},
  \bibinfo {author} {\bibfnamefont {D.}~\bibnamefont {Snoke}}, \bibinfo
  {author} {\bibfnamefont {L.}~\bibnamefont {Pfeiffer}}, \ and\ \bibinfo
  {author} {\bibfnamefont {K.}~\bibnamefont {West}},\ }\href {\doibase
  10.1126/science.1140990} {\bibfield  {journal} {\bibinfo  {journal}
  {Science}\ }\textbf {\bibinfo {volume} {316}},\ \bibinfo {pages} {1007}
  (\bibinfo {year} {2007})}\BibitemShut {NoStop}%
\bibitem [{\citenamefont {Cristofolini}\ \emph {et~al.}(2013)\citenamefont
  {Cristofolini}, \citenamefont {Dreismann}, \citenamefont {Christmann},
  \citenamefont {Franchetti}, \citenamefont {Berloff}, \citenamefont {Tsotsis},
  \citenamefont {Hatzopoulos}, \citenamefont {Savvidis},\ and\ \citenamefont
  {Baumberg}}]{Cristofolini2013prl}%
  \BibitemOpen
  \bibfield  {author} {\bibinfo {author} {\bibfnamefont {P.}~\bibnamefont
  {Cristofolini}}, \bibinfo {author} {\bibfnamefont {A.}~\bibnamefont
  {Dreismann}}, \bibinfo {author} {\bibfnamefont {G.}~\bibnamefont
  {Christmann}}, \bibinfo {author} {\bibfnamefont {G.}~\bibnamefont
  {Franchetti}}, \bibinfo {author} {\bibfnamefont {N.~G.}\ \bibnamefont
  {Berloff}}, \bibinfo {author} {\bibfnamefont {P.}~\bibnamefont {Tsotsis}},
  \bibinfo {author} {\bibfnamefont {Z.}~\bibnamefont {Hatzopoulos}}, \bibinfo
  {author} {\bibfnamefont {P.~G.}\ \bibnamefont {Savvidis}}, \ and\ \bibinfo
  {author} {\bibfnamefont {J.~J.}\ \bibnamefont {Baumberg}},\ }\href {\doibase
  10.1103/PhysRevLett.110.186403} {\bibfield  {journal} {\bibinfo  {journal}
  {Phys. Rev. Lett.}\ }\textbf {\bibinfo {volume} {110}},\ \bibinfo {pages}
  {186403} (\bibinfo {year} {2013})}\BibitemShut {NoStop}%
\bibitem [{\citenamefont {Tanese}\ \emph {et~al.}(2013)\citenamefont {Tanese},
  \citenamefont {Flayac}, \citenamefont {Solnyshkov}, \citenamefont {Amo},
  \citenamefont {Lema{\^\i}tre}, \citenamefont {Galopin}, \citenamefont
  {Braive}, \citenamefont {Senellart}, \citenamefont {Sagnes}, \citenamefont
  {Malpuech},\ and\ \citenamefont {Bloch}}]{Tanese2013ncomms}%
  \BibitemOpen
  \bibfield  {author} {\bibinfo {author} {\bibfnamefont {D.}~\bibnamefont
  {Tanese}}, \bibinfo {author} {\bibfnamefont {H.}~\bibnamefont {Flayac}},
  \bibinfo {author} {\bibfnamefont {D.}~\bibnamefont {Solnyshkov}}, \bibinfo
  {author} {\bibfnamefont {A.}~\bibnamefont {Amo}}, \bibinfo {author}
  {\bibfnamefont {A.}~\bibnamefont {Lema{\^\i}tre}}, \bibinfo {author}
  {\bibfnamefont {E.}~\bibnamefont {Galopin}}, \bibinfo {author} {\bibfnamefont
  {R.}~\bibnamefont {Braive}}, \bibinfo {author} {\bibfnamefont
  {P.}~\bibnamefont {Senellart}}, \bibinfo {author} {\bibfnamefont
  {I.}~\bibnamefont {Sagnes}}, \bibinfo {author} {\bibfnamefont
  {G.}~\bibnamefont {Malpuech}}, \ and\ \bibinfo {author} {\bibfnamefont
  {J.}~\bibnamefont {Bloch}},\ }\href {http://dx.doi.org/10.1038/ncomms2760}
  {\bibfield  {journal} {\bibinfo  {journal} {Nat. Commun.}\ }\textbf {\bibinfo
  {volume} {4}},\ \bibinfo {pages} {1749} (\bibinfo {year} {2013})}\BibitemShut
  {NoStop}%
\bibitem [{\citenamefont {Carusotto}\ and\ \citenamefont
  {Ciuti}(2004)}]{carusotto04prl}%
  \BibitemOpen
  \bibfield  {author} {\bibinfo {author} {\bibfnamefont {I.}~\bibnamefont
  {Carusotto}}\ and\ \bibinfo {author} {\bibfnamefont {C.}~\bibnamefont
  {Ciuti}},\ }\href {\doibase 10.1103/PhysRevLett.93.166401} {\bibfield
  {journal} {\bibinfo  {journal} {Phys. Rev. Lett.}\ }\textbf {\bibinfo
  {volume} {93}},\ \bibinfo {pages} {166401} (\bibinfo {year}
  {2004})}\BibitemShut {NoStop}%
\bibitem [{\citenamefont {Wouters}\ and\ \citenamefont
  {Carusotto}(2007)}]{wouters07pra}%
  \BibitemOpen
  \bibfield  {author} {\bibinfo {author} {\bibfnamefont {M.}~\bibnamefont
  {Wouters}}\ and\ \bibinfo {author} {\bibfnamefont {I.}~\bibnamefont
  {Carusotto}},\ }\href {http://link.aps.org/doi/10.1103/PhysRevA.76.043807}
  {\bibfield  {journal} {\bibinfo  {journal} {Phys. Rev. A}\ }\textbf {\bibinfo
  {volume} {76}},\ \bibinfo {pages} {043807} (\bibinfo {year}
  {2007})}\BibitemShut {NoStop}%
\bibitem [{\citenamefont {Keeling}\ and\ \citenamefont
  {Berloff}(2008)}]{keeling_spontaneous_2008}%
  \BibitemOpen
  \bibfield  {author} {\bibinfo {author} {\bibfnamefont {J.}~\bibnamefont
  {Keeling}}\ and\ \bibinfo {author} {\bibfnamefont {N.~G.}\ \bibnamefont
  {Berloff}},\ }\href {\doibase 10.1103/PhysRevLett.100.250401} {\bibfield
  {journal} {\bibinfo  {journal} {Phys. Rev. Lett.}\ }\textbf {\bibinfo
  {volume} {100}},\ \bibinfo {pages} {250401} (\bibinfo {year}
  {2008})}\BibitemShut {NoStop}%
\bibitem [{\citenamefont {Wouters}\ and\ \citenamefont
  {Savona}(2009)}]{Wouters2009prb}%
  \BibitemOpen
  \bibfield  {author} {\bibinfo {author} {\bibfnamefont {M.}~\bibnamefont
  {Wouters}}\ and\ \bibinfo {author} {\bibfnamefont {V.}~\bibnamefont
  {Savona}},\ }\href {\doibase 10.1103/PhysRevB.79.165302} {\bibfield
  {journal} {\bibinfo  {journal} {Phys. Rev. B}\ }\textbf {\bibinfo {volume}
  {79}},\ \bibinfo {pages} {165302} (\bibinfo {year} {2009})}\BibitemShut
  {NoStop}%
\bibitem [{\citenamefont {Lagoudakis}\ \emph {et~al.}(2011)\citenamefont
  {Lagoudakis}, \citenamefont {Manni}, \citenamefont {Pietka}, \citenamefont
  {Wouters}, \citenamefont {Liew}, \citenamefont {Savona}, \citenamefont
  {Kavokin}, \citenamefont {Andr{\'e}},\ and\ \citenamefont
  {{Deveaud-Pl{\'e}dran}}}]{lagoudakis11prlSHORT}%
  \BibitemOpen
  \bibfield  {author} {\bibinfo {author} {\bibfnamefont {K.~G.}\ \bibnamefont
  {Lagoudakis}}, \bibinfo {author} {\bibfnamefont {F.}~\bibnamefont {Manni}},
  \bibinfo {author} {\bibfnamefont {B.}~\bibnamefont {Pietka}}, \bibinfo
  {author} {\bibfnamefont {M.}~\bibnamefont {Wouters}}, \bibinfo {author}
  {\bibfnamefont {T.~C.~H.}\ \bibnamefont {Liew}}, \bibinfo {author}
  {\bibfnamefont {V.}~\bibnamefont {Savona}}, \bibinfo {author} {\bibfnamefont
  {A.~V.}\ \bibnamefont {Kavokin}}, \bibinfo {author} {\bibfnamefont
  {R.}~\bibnamefont {Andr{\'e}}}, \ and\ \bibinfo {author} {\bibfnamefont
  {B.}~\bibnamefont {{Deveaud-Pl{\'e}dran}}},\ }\href@noop {} {\bibfield
  {journal} {\bibinfo  {journal} {Phys. Rev. Lett.}\ }\textbf {\bibinfo
  {volume} {106}},\ \bibinfo {pages} {115301} (\bibinfo {year}
  {2011})}\BibitemShut {NoStop}%
\bibitem [{\citenamefont {Manni}\ \emph {et~al.}(2012)\citenamefont {Manni},
  \citenamefont {Lagoudakis}, \citenamefont {Liew}, \citenamefont {Andr\'e},
  \citenamefont {Savona},\ and\ \citenamefont {Deveaud}}]{Manni2012natcomm}%
  \BibitemOpen
  \bibfield  {author} {\bibinfo {author} {\bibfnamefont {F.}~\bibnamefont
  {Manni}}, \bibinfo {author} {\bibfnamefont {K.~G.}\ \bibnamefont
  {Lagoudakis}}, \bibinfo {author} {\bibfnamefont {T.~C.~H.}\ \bibnamefont
  {Liew}}, \bibinfo {author} {\bibfnamefont {R.}~\bibnamefont {Andr\'e}},
  \bibinfo {author} {\bibfnamefont {V.}~\bibnamefont {Savona}}, \ and\ \bibinfo
  {author} {\bibfnamefont {B.}~\bibnamefont {Deveaud}},\ }\href
  {http://dx.doi.org/10.1038/ncomms2310} {\bibfield  {journal} {\bibinfo
  {journal} {Nat. Commun.}\ }\textbf {\bibinfo {volume} {3}},\ \bibinfo {pages}
  {1309} (\bibinfo {year} {2012})}\BibitemShut {NoStop}%
\bibitem [{\citenamefont {Vi{\~n}a}\ \emph {et~al.}(2004)\citenamefont
  {Vi{\~n}a}, \citenamefont {Andr{\'e}}, \citenamefont {Ciulin}, \citenamefont
  {Ganiere},\ and\ \citenamefont {Deveaud}}]{vina2004sst}%
  \BibitemOpen
  \bibfield  {author} {\bibinfo {author} {\bibfnamefont {L.}~\bibnamefont
  {Vi{\~n}a}}, \bibinfo {author} {\bibfnamefont {R.}~\bibnamefont {Andr{\'e}}},
  \bibinfo {author} {\bibfnamefont {V.}~\bibnamefont {Ciulin}}, \bibinfo
  {author} {\bibfnamefont {J.~D.}\ \bibnamefont {Ganiere}}, \ and\ \bibinfo
  {author} {\bibfnamefont {B.}~\bibnamefont {Deveaud}},\ }\href
  {http://stacks.iop.org/0268-1242/19/i=4/a=110} {\bibfield  {journal}
  {\bibinfo  {journal} {Semicond. Sci. Technol.}\ }\textbf {\bibinfo {volume}
  {19}},\ \bibinfo {pages} {S333} (\bibinfo {year} {2004})}\BibitemShut
  {NoStop}%
\bibitem [{\citenamefont {Shelykh}\ \emph {et~al.}(2005)\citenamefont
  {Shelykh}, \citenamefont {Vi{\~n}a}, \citenamefont {Kavokin}, \citenamefont
  {Galkin}, \citenamefont {Malpuech},\ and\ \citenamefont
  {Andr{\'e}}}]{Shelykh2005ssc}%
  \BibitemOpen
  \bibfield  {author} {\bibinfo {author} {\bibfnamefont {I.~A.}\ \bibnamefont
  {Shelykh}}, \bibinfo {author} {\bibfnamefont {L.}~\bibnamefont {Vi{\~n}a}},
  \bibinfo {author} {\bibfnamefont {A.~V.}\ \bibnamefont {Kavokin}}, \bibinfo
  {author} {\bibfnamefont {N.~G.}\ \bibnamefont {Galkin}}, \bibinfo {author}
  {\bibfnamefont {G.}~\bibnamefont {Malpuech}}, \ and\ \bibinfo {author}
  {\bibfnamefont {R.}~\bibnamefont {Andr{\'e}}},\ }\href {\doibase
  http://dx.doi.org/10.1016/j.ssc.2005.04.012} {\bibfield  {journal} {\bibinfo
  {journal} {Solid State Commun.}\ }\textbf {\bibinfo {volume} {135}},\
  \bibinfo {pages} {1} (\bibinfo {year} {2005})}\BibitemShut {NoStop}%
\bibitem [{\citenamefont {Read}\ \emph {et~al.}(2009)\citenamefont {Read},
  \citenamefont {Liew}, \citenamefont {Rubo},\ and\ \citenamefont
  {Kavokin}}]{Read2009prb}%
  \BibitemOpen
  \bibfield  {author} {\bibinfo {author} {\bibfnamefont {D.}~\bibnamefont
  {Read}}, \bibinfo {author} {\bibfnamefont {T.~C.~H.}\ \bibnamefont {Liew}},
  \bibinfo {author} {\bibfnamefont {Y.~G.}\ \bibnamefont {Rubo}}, \ and\
  \bibinfo {author} {\bibfnamefont {A.~V.}\ \bibnamefont {Kavokin}},\ }\href
  {\doibase 10.1103/PhysRevB.80.195309} {\bibfield  {journal} {\bibinfo
  {journal} {Phys. Rev. B}\ }\textbf {\bibinfo {volume} {80}},\ \bibinfo
  {pages} {195309} (\bibinfo {year} {2009})}\BibitemShut {NoStop}%
\bibitem [{\citenamefont {Wouters}\ \emph {et~al.}(2010)\citenamefont
  {Wouters}, \citenamefont {Liew},\ and\ \citenamefont
  {Savona}}]{Wouters2010prb}%
  \BibitemOpen
  \bibfield  {author} {\bibinfo {author} {\bibfnamefont {M.}~\bibnamefont
  {Wouters}}, \bibinfo {author} {\bibfnamefont {T.~C.~H.}\ \bibnamefont
  {Liew}}, \ and\ \bibinfo {author} {\bibfnamefont {V.}~\bibnamefont
  {Savona}},\ }\href {\doibase 10.1103/PhysRevB.82.245315} {\bibfield
  {journal} {\bibinfo  {journal} {Phys. Rev. B}\ }\textbf {\bibinfo {volume}
  {82}},\ \bibinfo {pages} {245315} (\bibinfo {year} {2010})}\BibitemShut
  {NoStop}%
\bibitem [{\citenamefont {Wouters}(2012)}]{Wouters2012njp}%
  \BibitemOpen
  \bibfield  {author} {\bibinfo {author} {\bibfnamefont {M.}~\bibnamefont
  {Wouters}},\ }\href {http://stacks.iop.org/1367-2630/14/i=7/a=075020}
  {\bibfield  {journal} {\bibinfo  {journal} {New J. Phys.}\ }\textbf {\bibinfo
  {volume} {14}},\ \bibinfo {pages} {075020} (\bibinfo {year}
  {2012})}\BibitemShut {NoStop}%
\bibitem [{\citenamefont {{Solnyshkov}}\ \emph {et~al.}()\citenamefont
  {{Solnyshkov}}, \citenamefont {{Ter{\c c}as}}, \citenamefont {{Dini}},\ and\
  \citenamefont {{Malpuech}}}]{Solnyshkov2013arXiv}%
  \BibitemOpen
  \bibfield  {author} {\bibinfo {author} {\bibfnamefont {D.~D.}\ \bibnamefont
  {{Solnyshkov}}}, \bibinfo {author} {\bibfnamefont {H.}~\bibnamefont {{Ter{\c
  c}as}}}, \bibinfo {author} {\bibfnamefont {K.}~\bibnamefont {{Dini}}}, \ and\
  \bibinfo {author} {\bibfnamefont {G.}~\bibnamefont {{Malpuech}}},\
  }\href {\doibase 10.1103/PhysRevA.89.033626} {\bibfield
  {journal} {\bibinfo  {journal} {Phys. Rev. A}\ }\textbf {\bibinfo {volume}
  {89}},\ \bibinfo {pages} {033626} (\bibinfo {year} {2014})} \BibitemShut {NoStop}%
\bibitem [{\citenamefont {{Sieberer}}\ \emph {et~al.}()\citenamefont
  {{Sieberer}}, \citenamefont {{Huber}}, \citenamefont {{Altman}},\ and\
  \citenamefont {{Diehl}}}]{Sieberer2013arxive}%
  \BibitemOpen
  \bibfield  {author} {\bibinfo {author} {\bibfnamefont {L.~M.}\ \bibnamefont
  {{Sieberer}}}, \bibinfo {author} {\bibfnamefont {S.~D.}\ \bibnamefont
  {{Huber}}}, \bibinfo {author} {\bibfnamefont {E.}~\bibnamefont {{Altman}}}, \
  and\ \bibinfo {author} {\bibfnamefont {S.}~\bibnamefont {{Diehl}}},\
  }\href {\doibase 10.1103/PhysRevB.89.134310} {\bibfield
  {journal} {\bibinfo  {journal} {Phys. Rev. B}\ }\textbf {\bibinfo {volume}
  {89}},\ \bibinfo {pages} {134310} (\bibinfo {year} {2014})} \BibitemShut {NoStop}%
\bibitem [{\citenamefont {Rojas}(1996)}]{rojas96book_neural}%
  \BibitemOpen
  \bibfield  {author} {\bibinfo {author} {\bibfnamefont {R.}~\bibnamefont
  {Rojas}},\ }\href {http://books.google.es/books?id=txsjjYzFJS4C} {\emph
  {\bibinfo {title} {Neural Networks}}}\ (\bibinfo  {publisher}
  {Springer-Verlag, New York},\ \bibinfo {year} {1996})\BibitemShut {NoStop}%
\end{thebibliography}
\end{document}